%%%%%%%%%%%%%%%%%%%%%%%%%%%%%%%%%%%%%%%%%%%%%%%%%%%%%%%%%%%%%%%%%%%%%%%%%%%
%%%%%%%%%%%%%%%%%%%
\input phyzzx
\hfuzz 11pt
\font\mybb=msbm10 at 12pt

\def\Bbb#1{\hbox{\mybb#1}}

\def\bE{\Bbb {E}}

\def\bfomega{\omega\kern-7.0pt \omega}

%%%%%%%%%%%%%%%%%%%%%%%%%%%%%%%%%%%%%%%%%%%%%%%%%%%%%%%%%%%%%%%%%%%%

\REF\kshir { K Shiraishi, {\sl Nucl. Phys.} {\bf B402} (1993) 399-410.}
\REF\gibrub {G W Gibbons \& P J Ruback, {\sl Phys. Rev. Lett.} {\bf 57}
 (1986) 1492.}
\REF\rub{ P J Ruback, {\sl Commun. Math. Phys. } {\bf 107} (1986) 93.}
\REF\ferear{ R C Ferrell \& D M Eardley, {\sl Phys. Rev. Lett.} 
{\bf 59} (1987) 1617.}   
\REF\felsam{ A G Felce \& T M Samols, {\sl Phys. Letts.} {\bf B308} 
(1993) 30: hep-th/921118.}
\REF\gppkt{G. Papadopoulos \& P.K. Townsend, {\sl Phys. Lett.}
 {\bf B380} (1996) 273.}
\REF\gibkal{ G. W. Gibbons \& R. Kallosh, {\sl Phys. Rev. }{\bf D51}
 (1995) 2839.}
\REF\howepap{P.S. Howe \& G. Papadopoulos, {\sl Nucl. Phys.} {\bf B289}
(1987) 264; {\sl Class. Quantum Grav.} {\bf 5} (1988) 1647.}
\REF\hullone{C.M. Hull, {\sl Lectures on Nonlinear 
Sigma Models and Strings},
 Lectures given in the {\sl Super Field Theories} workshop, 
Vancouver Canada (1986),
published in Vancouver Theory Workshop.}
\REF\hitchin {N.J. Hitchin, A. Karlhede, U. Lindstr\"om \& M. Ro\v cek, 
{\sl Commun. Math. Phys.}{\bf 108} (1987) 535. }
\REF\howepapd{P.S. Howe \& G. Papadopoulos, {\sl Phys. Lett. }
 {\bf B379} (1996) 80.}
\REF\harveyoct{J. Harvey \& A. Strominger, {\sl Phys. Rev. Let.}
 {\bf 5} (1991) 549.}
\REF\ivan{T.A.  Ivanova, {\sl Phys. Lett.} {\bf B315} (1993) 277.} 
\REF\nicolai{ M. G\"unaydin \& H. Nicolai, {\sl Phys. Lett.}
 {\bf B351} (1995) 169: hep-th/9502009;
Addendum-ibid {\bf B376} (1996) 329.}
\REF\duffoct{M.J. Duff, J.M. Evans, R.R. Khuri, J.X. Lu, \&
R. Minasian, {\sl The Octonionic Membrane} hep-th/9706124.}
\REF\zumino{B. Zumino, {\sl Phys. Lett.} {\bf B87} (1979) 203.}
\REF\gates{S.J. Gates, C.M. Hull \& M. Ro\v cek, {\sl Nucl. Phys.} 
{\bf B248} (1984) 157.}
\REF\coles{ R. Coles \& G. Papadopoulos, {\sl Class. Quantum Grav.}
 {\bf 7} (1990) 427.}
\REF\gibbons{G. Gibbons, {\sl Nucl. Phys.} {\bf B207} (1982) 337.}
\REF\tseytlin{A.A. Tseytlin, {\sl Nucl. Phys.} {\bf B475} (1996) 149:
 hep-th/9604035.}
\REF\kastor{J.P. Gauntlett, D.A. Kastor \& J. Traschen, 
{\sl Nucl. Phys. } {\bf B478} (1996) 544:
hep-th/9604179.}
\REF\tseytlinb{A.A. Tseytlin, {\sl Mod. Phys. Lett.} {\bf A111} (1996) 689.}
\REF\manton{G.W. Gibbons \& N. Manton, {\sl Phys. Lett.} {\bf B356}
 (1995) 32: hep-th/9506052.}
\REF\gaunhar{ J. P. Gauntlett, J. A. Harvey, M. M. Robinson \&
D. Waldram {\sl Nucl Phys } {\bf B411}
(1994) 461.}
\REF\duff {M.J. Duff \& J.X.  Lu, {\sl Nucl. Phys.} {\bf B416} (1994) 301. }
\REF\papad{G. Papadopoulos, {\it The Universality of M-branes}, Talk given at
the Imperial College Workshop on {\it Gauge Theories, Applied Supersymmetry and
Quantum Gravity} (1996), hep-th/9611029.}
\REF\howepapc{P.S. Howe \& G. Papadopoulos, {\sl Commun. Math. Phys.} {\bf 151}
(1993) 467.}
\REF\calhar{ C. G. Callan, J. A. Harvey \& A. Strominger, 
{\sl Nucl. Phys. } {\bf B359} (1991) 611.} 
\REF\aurreg{  R. D'Auria \& T. Regge, {\sl Nucl. Phys. } {\bf B195}
(1982) 308.}
\REF\rey{ S. J. Rey, {\sl Phys. Rev.} {\bf D43} (1991) 526.} 
\REF\rham{ J. Rhamfeld, {\sl Phys. Lett.} {\bf B372} (1996) 198.} 
\REF\gibvan{G.W. Gibbons, R.H. Rietdijk \& J.W. van Holten, 
{\sl Nucl. Phys.} {\bf B404} (1993) 42.}
\REF\howepapb{P.S. Howe \& G. Papadopoulos, {\sl Nucl .Phys.} 
{\bf B381} (1992) 360. }
\REF\rocek{M. Ro\v cek, K. Schoutens \& A. Sevrin, {\sl Phys. Lett.}
 {\bf B265} (1991) 303.}
\REF\sfetsos{F. de Jonghe, K. Peeters \& K. Sfetsos, 
{\sl Class. Quantum Grav.} {\bf 14} (1997) 35.}
\REF\duffstelle{M.J. Duff \& K.S. Stelle, {\sl Phys. Lett.} {\bf B253}
 (1991) 113.}
\REF\dghrr{A. Dabholkar, G.W. Gibbons, J.A. Harvey \& F. Ruiz-Ruiz, 
{\sl Nucl. Phys.} {\bf B340} (1990) 33.}
\REF\boundstates{
N. Khviengia, Z. Khviengia, H. L\"u \& C.N. Pope, {\sl Phys. Lett.}
 {\bf B388} (1996) 21;\hfil\break
M.J. Duff \& J. Rahmfeld, {\sl Nucl. Phys.} {\bf B481} (1996) 332.}
\REF\buscher{ T. Buscher,{ \sl  Phys. Lett.}{\bf B201} (1988) 466; 
{\bf B194} (1987) 59. }
\REF\bakas{ I. Bakas \& K. Sfetsos, {\sl Phys. Lett.} {\bf B349} (1995) 448.}
\REF\lust{E. Kiritsis, C. Counnas \&  D. L\"ust {\sl Int. Journ.
 Mod. Phys.} {\bf A9} (1994) 1361.}
\REF\town{J.P. Gauntlett, G.W. Gibbons, G. Papadopoulos \&
P.K. Townsend, {\sl Hyper-Kahler
manifolds and multiply intersecting branes} Nucl. Phys. (1997),
 to appear: hep-th/9702202.}
\REF\douglas {M. R. Douglas, D. Kabat, P. Pouliot \& S.H. Shenker,
 {\it D-branes and Short Distances in String Theory} hep-th/9608024.}
\REF\douglaspolch {M. Douglas, J. Polchinski \& A. Strominger, 
{\it Probing Five-Dimensional Black Holes with D-branes}, hep-th/9703031.}
 \REF\bonneau{ G. Bonneau \& G. Valent, {\sl Local heterotic
 geometry in holomorphic
co-ordinates}, hep-th/9401003.}
\REF\papadb{G. Papadopoulos, {\sl Phys. Lett.} {\bf B356} (1995) 249. }
\REF\gibhawk{G.W. Gibbons \& S.W. Hawking, {\sl Phys. Lett.} 
{\bf 78B} (1978) 430.}
\REF\gibruba{G.W. Gibbons \& P. J. Ruback, {\sl Commun. Math. Phys.} 
{\bf 115} (1988) 267.}
\REF\callanm{ C.G. Callan \& J.  Maldacena, {\sl Nucl. Phys.}
{\bf B472} (1996) 591: hep-th/9602043.}
\REF\modfive{D.M. Kaplan and J. Michelson, {\sl Scattering of several
multiply charged extremal D=5 black holes}, hep-th/9707021.}
\REF\geod{V. Benci \& F. Giannoni, {\sl Duke Math. Journ.}
{\bf 68} (1992) 195.}
\REF\das{S.R. Das, G.W. Gibbons \& S.D. Mathur, 
{\sl Phys. Rev. Letts.} {\bf 78} (1997) 417: hep-th/9609052.}
\REF\cornish{ N. Cornish \& G.W. Gibbons, 
{\sl Class. Quantum Grav.} {\bf 14} (1997) 1865: gr-qc/9612060.}
%%%%%%%%%%%%%%%%%%%%%%%%%%%%TITLE PAGE%%%%%%%%%%%%%%%%%%%%%%%%%
\Pubnum{ \vbox{ \hbox{R/97/28}\hbox{Imperial/TP/96-97/50} } }
\pubtype{}
\date{June, 1997}
\titlepage
\title{HKT and OKT Geometries on Soliton Black Hole Moduli Spaces}
\author{G.W. Gibbons, G. Papadopoulos}
\address{DAMTP, Silver Street, University of Cambridge, Cambridge CB3 9EW}
\andauthor{K.S. Stelle}
\address{The Blackett Laboratory, Imperial College, Prince Consort Road,
London SW7 2BZ}
%\address{}
\abstract {We consider Shiraishi's metrics on the moduli space of extreme
black holes. We interpret the simplification in the pattern of N-body
interactions that he observed in terms of the recent picture of black holes in
four and five dimensions as composites, made up of intersecting branes.  We
then show that the geometry of the moduli space of a class of
black holes in five and nine dimensions is hyper-K\"ahler with torsion, and
octonionic-K\"ahler with torsion, respectively. For this, we examine the
geometry of point particle models with extended world-line supersymmetry and
show that both of the above geometries arise naturally in this context. In
addition, we construct a large class of hyper-K\"ahler with torsion and
octonionic-K\"ahler with torsion geometries in various dimensions.   We also
present a brane interpretation of our results. }

\endpage
\pagenumber=2

%%%%%%%%%%%%%%%%%%%%%%%%%%%%%%%%%%%%%%%%%%%%%%%%%%%%%%%%%%%%%%%%%%

%\def\I{\rlap I\mkern3mu{\rm I}}

\def\C{\mkern1mu\raise2.2pt\hbox{$\scriptscriptstyle|$}\mkern-7mu{\rm C}}

\def\exp{{\rm exp}}
\def\log{{\rm log}}

%\sequentialequations
\def\ft#1#2{{\textstyle{{\scriptstyle #1}\over {\scriptstyle #2}}}}

%%%%%%%%%%%%%%%%%%%%%%%%%%%%%%%%%%%%%%%%%%%%%%%%%%%%%%%%%%%%%%%%%%%%%%%

\chapter{ Introduction}

Recent years have seen great progress in our understanding of
string-theory and M-theory by considering the non-perturbative effects
of classical solutions of the associated low energy supergravity
theories representing $p$-branes.  Supersymmetric ({\it i.e.}\ BPS)
solutions describing $k$ parallel or intersecting $p$-branes typically
depend on a number of harmonic functions on an $n$-dimensional
Euclidean or conformally Euclidean transverse space. In the simplest
case of just one harmonic function $H$ on ${\Bbb E}^n$, $H$ is taken
to be one plus a sum of simple isolated poles located at positions
${\bf x}_i \in {\Bbb E}^n$.  
$$ 
H= 1 + \sum_{i=1} ^k { \mu_i \over {
(n-2) |{\bf x} - {\bf x}_i |^{n-2} } } \eqn\intone 
$$ 
The residues, $\mu \over n-2$ , of the poles are fixed by quantization
conditions but the locations may be freely specified and so, if the
$p$-branes are distinct, the moduli space ${\cal M}_k$ of such
solutions is the configuration space of $k$ particles moving on ${\Bbb
E}^n$, that is $ {\tilde C} _k ({\Bbb E}^n ) \equiv \bigl ({\Bbb
E}^n\bigr )^k \\ \Delta$ where $\Delta$ is the diagonal set when two
or more positions coincide. If the $p$-branes are identical, one
should quotient by $S_k$, the permutation group on $k$ letters, to
obtain $ C_k ({\Bbb E}^n )={\tilde C} _k ({\Bbb E}^n )/S_k$, but for
the time being we shall ignore this point and shall always work on the
covering space ${\tilde C} _k ({\Bbb E}^n )$. If $n>2$, which is the
case we are mainly interested in, this covering space is simply
connected. If more than one harmonic function is involved, the moduli
space will still be ${\tilde C} _k ({\Bbb E}^n )$ but now one must
include the poles of all the harmonic functions; $k$ in this case is
the number of different poles of all harmonic functions.

 The metrics on the moduli spaces associated with extreme black holes have
been known for sometime [\kshir, \gibrub, \rub, \ferear, \felsam]
 but as yet their geometric significance has been obscure. The
interactions of the black holes depend on the `dilaton' coupling $a$
(whose precise definition is in the next section). One of Shiraishi's
observations was that although in general there are N-body
interactions with arbitrary N, these simplify in the special cases
$a=0,1/{\sqrt {3}}, 1, {\sqrt 3}$ in four dimensions, and in the
special cases $a=0,1,2$ in five dimensions.  Moreover in the case
$a=1$ in all dimensions the metric dramatically simplifies and gives
rise to just two-body interactions.  In this paper, we give an
interpretation of these observations using the recent picture of black
holes as composites made up of intersecting branes [\gppkt].  The
simplification in the $a=1$ case arises because the solutions preserve
$1/4$ of maximal (eleven-dimensional) rigid supersymmetry.  Another
result of this paper is to show that the moduli space of
five-dimensional ($n=4$) $a=1$ BPS black holes is a hyper-K\"ahler
manifold with torsion (HKT) and that the moduli space of a class of
nine-dimensional ($n=8$) $a=1$ BPS black holes is an
octonionic-K\"ahler manifold with torsion (OKT). The relevance of 
torsion in black hole moduli spaces was first pointed out in [\gibkal].

The HKT geometry has been found in the context of two-dimensional (4,0)
supersymmetric sigma models and it is a generalisation of the hyper-K\"ahler
geometry [\hullone, \howepap].  There is a close relation between hyper-K\"ahler and HKT
geometries. For example, both admit a twistor construction and they can be
reconstructed from data on their twistor spaces [\hitchin,\howepapd].

The OKT geometry will be found in the context of one-dimensional N=8
supersymmetric sigma models, {\it i.e.}\ of N=8 supersymmetric particle
mechanics.  The underlying algebraic structure of this geometry is that of the
octonions.  There is some similarity of this geometry to that of the
octonionic string [\harveyoct, \ivan, \nicolai] and octonionic membrane [\duffoct]
solutions of supergravity theories but we have not managed to establish a
direct relation.

Our investigation of the geometry of the moduli space of black holes is guided
by the relation between the number of supersymmetries of a sigma model and the
geometry of its target space [\zumino, \gates, \hullone, \howepap].  For this we shall
summarize the results of [\coles] on the geometry of one-dimensional N=1
supersymmetric sigma models. Then we shall examine the target space geometry
of one-dimensional sigma
models with extended supersymmetry generalizing some of the conditions 
found in [\coles].  In
addition, we shall identify the geometry of the target space of a class of 
one-dimensional sigma models
with eight supersymmetries as that of OKT geometry.

We shall also present a brane interpretation of our results.  Following
[\gppkt],  we shall show that the black-hole solutions that we investigate
have a ten-dimensional interpretation.  In particular, the five-dimensional
black hole solution can be lifted to the ten-dimensional solution of IIB
supergravity having the interpretation of two three branes intersecting on a
string, and the nine-dimensional black hole can be lifted to the
ten-dimensional solution of IIA supergravity having the interpretation of a
wave on a string.  Then, using the ten-dimensional solutions, we shall specify
the one-dimensional supersymmetry multiplet that describes the sector of the
effective theory of the black holes that is related to the geometry of the
associated moduli spaces.

This paper is organised as follows: in section two, we summarise the results
of Shiraishi on the metrics on higher-dimensional black hole moduli spaces and
then we explain some of his conclusions, using the fact that some of black hole
solutions can be thought of as composite objects. In section 3, we investigate
the various geometries that arise in the context of one-dimensional
supersymmetric sigma models and define the OKT geometry. In section 4, we
construct a number of examples of HKT geometries and show that the moduli space
of a class of five-dimensional black holes is an HKT manifold. In section 5, we
construct a number of examples of OKT geometries and show that the moduli space
of a class of eight-dimensional black holes is an OKT manifold. In section 6,
we use the interpretation of these black holes as intersecting-brane solutions
of ten-dimensional supergravity theories to determine the nature of their
effective theory. In section 7, we examine some of the brane probe geometries
that arise in the context of five-dimensional black holes.  In section 8,
we remark on the structure of the geodesics of the HKT and OKT geometries that
we have found and discuss the quantum behaviour of these metrics. In addition,
we comment on the  moduli spaces of other black hole solutions in various
dimensions. Finally we include an appendix in which we present some results on
harmonic forms and harmonic spinors on our moduli spaces.

\chapter{ Shiraishi metrics}

The slow motion of parallel $p$-branes is expected
to give rise to a classical particle  motion on ${\cal M}_k$ 
and, quantum mechanically, to the quantization of that classical system.
In a supersymmetric theory  the classical system
is extended to that of a supersymmetric spinning particle.
The simplest action for the bosonic part of the system
is based on geodesic motion with respect to an 
appropriate Riemannian metric $g_{\alpha \beta}$ on ${\cal M}_k$, 
but more elaborate possibilities could also be envisaged. 

Rather than consider the motion of $p$-branes in $p+1+n$ spacetime
dimensions one may, in view of the invariance of the setup under the
action of ${\Bbb R}^p$ acting as translations on the p-brane world
volume, dimensionally reduce and consider the equivalent problem of
particles, {\it i.e.}\ $0-$branes, moving in $n+1$ spacetime
dimensions.  In the case that only one harmonic function is involved
one may, possibly by making suitable duality transformations, regard
the particles as carrying an electric charge associated to an abelian
2-form $F$ and also interacting via the exchange of a massless scalar
$\phi$ . If more than one harmonic function is involved then, in some
cases, one needs to consider more than one 2-from and more than one
scalar.

In the simple case of one harmonic function the equivalent Lagrangian in 
$n+1$ dimensions is
$$
R- {4 \over n-1} (\nabla \phi )^2 - e^ {-{ 4a \over n-1}\phi} F^2
\eqn\inttwo
$$
The solutions are given by
$$
ds^2= - H^{ - { 2(n-2) \over n-2+a^2}}  dt^2+
  H^{2 \over n-2+a^2} d{\bf x}^2 
\eqn\intthree   
$$  
$$
F= \pm \sqrt { n-1 \over 2(n-2+a^2) }d \bigl ( {dt \over H }\bigr )\ ,
\eqn\intfour
$$
and
$$
e^{- {4a \over n-1} \phi } = H^{ 2 a^2 \over n-2+a^2}\ .
\eqn\intfive
$$

The dimensionless constant $a$ depends upon precisely
what objects are being
considered. For the moment we leave it unspecified. Note that

\item {(i)}  The correspondence between $p-$branes in $p+1+n$ dimensions and
0-branes in $n+1$ dimensions is not one-one. For example two different 
$p-$branes may reduce to give the same solution in $n+1$ spacetime dimensions.

\item {(ii)}  The solutions in $n+1$ dimensions
are in general singular. Only the case $a=0$ corresponds to
a regular (Reissner-Nordstr\"om-Tangherlini) black hole
with finite event horizon area.
In general a  solution in higher dimensions,
singular or not, 
will reduce, in most cases, to a singular solution in $n+1$ spacetime
 dimensions.
However, one expects that non-singular 
solutions in $n+1$ spacetime dimensions
will lift to non-singular solutions in higher dimensions.

One may now compute the metric on ${\cal M}_k$ directly from the
classical theory.  This programme was initiated in [\gibrub] by
calculating the asymptotic metrics at large separation in the case
$n=3$. The exact metric was calculated in the case $a=0$ and $n=3$ in
[\ferear] and the exact metrics for all $a$ and $n$ were worked out by
Shiraishi [\kshir] in terms of an integral over ${\Bbb R}^n$.  His
result, which includes all previous ones as special cases, is
$$
\eqalign{
ds^2 &=\sum_{i=1}^k  m_i d{\bf x}_i^2 
+ { (n-1) (n-a^2) \over 8 \pi ( n-2+a^2)} 
\times 
\cr&\int d^n{\bf x}\, H^{ 2 (1-a^2) \over n-2+a^2}(x)\, \sum_{i<j} ^k 
{ ( {\bf x } -{\bf x}_i )   .  ( {\bf x} - {\bf x}_j )   
|d{\bf x}_i -d {\bf x}_j|^2 \mu_i \mu _j 
\over 
|{\bf x} - {\bf x}_i |^n  | {\bf x} -{\bf x} _j | ^n}\ ,}
\eqn\aone   
$$
where the mass $m_i$ of i'th particle is given by
$$
m_i = { \pi ^{ {n \over 2} -1} (n-1) \over  4 ( n-2+ a^2 ) 
\Gamma ( { n \over 2}) } \, \mu _i\ , 
\eqn\atwo
$$
and $H$ is given in \intone.
The normalization of the metric is determined by the condition that
the kinetic energy is ${ 1\over 2} g_{\alpha \beta} {\dot x}^\alpha 
{\dot x} ^ \beta$.

Shiraishi noticed a number of striking  features of his metrics,
some of which had been noticed previously in special cases.

\item{\bullet} In general the geodesic motion arises from
 $N-$body velocity-dependent forces for all values of $N$.
However in special cases the forces simplify.

\item{\bullet} If $a=0$ and $n=3$, there are only $2,3$ and $4$ body forces.

\item{\bullet} If $a=0$ and $n=4$, there are only $2$ and $3$ body forces.

\item {\bullet} If $a= { 1\over \sqrt3}$ and $n=3$, there are only $2$ and $3$
body forces.

\item{\bullet} If $a=1$ for all $n$, there are only 2-body forces. The
asymptotic metric is exact in this case. 

\item{ \bullet } If $a^2= n$  then the metric is flat and the asymptotic metric
is also exact in this case.

In retrospect these observations may be understood as follows. 

In {\sl four} spacetime dimensions there is a family of regular 
black hole solutions depending upon four independent harmonic functions
$(H_1,H_2,H_3, H_4)$. These black-hole solutions  can be lifted to solutions
of eleven-dimensional  supergravity which have the interpretation of
intersecting branes preserving $1/8$ of the spacetime supersymmetry [\gppkt,
\tseytlin]. Each harmonic function is associated with a brane involved in the
intersection [\tseytlin, \kastor].  We have the following specializations and numbers of
supersymmetries when only one harmonic function is involved:

\item {(i)} $a=0 \equiv (H,H,H,H) \leftrightarrow 4$.

\item {(ii)} $a={ 1\over {\sqrt 3} } \equiv (1,H,H,H) \leftrightarrow 4$

\item {(iii)} $a= 1 \equiv (1,1,H,H) \leftrightarrow  8$.

\item {(iv)} $a= {\sqrt 3} \equiv (1,1,1,H) \leftrightarrow  16$.

The N-body structure of the velocity dependent forces between the
black
 holes is a reflection of
their composite nature.  It is rather striking that this reveals
itself
 in this way.  Put another
way, by scattering black holes against one-another, one could in
 principle unravel
`experimentally' their composite nature, and learn for example
 that if $a=1/{\sqrt 3}$ then three
basic objects are involved.

In {\sl five}  spacetime dimensions there is a family of regular 
black hole solutions depending upon three independent harmonic functions
$(H_1,H_2,H_3)$ (see [\tseytlinb]).  These black hole solutions  again can be lifted
 to solutions of
eleven-dimensional supergravity preserving $1/8$ of the spacetime
 supersymmetry which have the
interpretation of intersecting branes [\gppkt, \tseytlin ]. Each
 harmonic function is associated with
a brane involved in the intersection. We have the following
 specializations and numbers of
supersymmetries when only one harmonic function is involved:

\item {(i)} $a=0 \equiv (H,H,H) \leftrightarrow 4$

\item {(ii)} $ a= 1 \equiv (1,H,H) \leftrightarrow 8$

\item {(iii)} $a= {\sqrt 3} \equiv (1,1,H) \leftrightarrow 16$

The case $a^2= n$ arises if one dimensionally reduces a vacuum pp-wave
from $n+2$ spacetime dimensions, the electric charge arising as a
Kaluza-Klein charge. The case $n=3$ was noted in [\gibbons ] and the
electromagnetically dual case (Kaluza-Klein monopoles) directly
verified to be flat in [\rub ]. The case $n= 4$ corresponds to the
motion of parallel five branes and was directly shown to be flat in
[\felsam]. The case $n=9$ is the $D-$particle of type IIA string
theory.  It is clear that in all these cases the flatness of the
moduli spaces arises as a consequence of the high degree of supersymmetry,
{\it i.e.}\ 16.

One should be able to understand these metrics entirely
from the point 
of view of supersymmetry. Naively one might have anticipated
 the following correspondences between the number of supersymmetries preserved
by a solution and the geometry of its moduli space 

\item {(i)} $4 \equiv$   Complex or Hyper-complex  Geometry.

\item {(ii)} $8 \equiv$   Hyper-complex or \lq\lq Octonionic" Geometry.

\item {(iii)} $16 \equiv$   Flat Geometry.

However this immediately raises some puzzles because 
of the dimensionality, $nk$,  of the moduli spaces.
This may indicate that there are some \lq\lq missing" moduli
which have not been taken into account.

Let us consider the case $a^2=1$. The metric dramatically simplifies
to give
$$
ds^2 = \sum _ i m_i d{\bf x}_i .d{\bf  x}_i + 
{\pi ^ { {n\over 2}-1} \over 4( n-2)  \Gamma ({n \over 2})   }  
  \sum ^k _{i<j} 
{|d {\bf x}_i - d {\bf x}_j |^2 \mu_i \mu _j
\over  |{\bf x}_i - {\bf x}_j |^{ n-2}}\ .
\eqn\athree
$$

It has already been observed [\manton ] that if $n=3$ the Shiraishi
metric
 coincides
with that on the quotient ${\cal N}_{4k} /  T^k$ of the Gibbons-Manton
$4k$-dimensional Hyper-K\"ahler manifold ${\cal N}_{4k}$,  admitting a
triholomorphic $T^k$ action, where $T^k$ is the $k$-dimensional torus group.
The Gibbons-Manton metric which is relevant for the black-hole moduli spaces
has the opposite sign for the \lq\lq mass parameters" from that 
relevant for BPS monopoles. The fact that the metric is hyper-K\"ahler  is
consistent with $(4,4)$ supersymmetry (in the two-dimensional sense) but
confirms the existence of missing moduli which have not yet been precisely
identified. We remark that if $a^2=1$ and $n=8$, then the metric on the moduli
space of two black holes is similar to that one obtains for the metric on
the moduli space  of two heterotic strings studied  by Gauntlett et al
[\gaunhar]. However, the latter metric  has 
 \lq\lq mass parameters" with the opposite sign from those of the
former. The effective theory associated with the heterotic strings has $(8,0)$
supersymmetry (in the two-dimensional sense).

The main observations of the present paper concern the cases $a=1$, $n=4$ and
$a=1$, $n=8$. If the $a=1$, $n=4$ solution is lifted up to six spacetime
dimensions, it becomes the completely non-singular self-dual string solution
[\duff]. If it further is lifted to ten dimensions, it becomes the solution of
IIB supergravity with two 3-branes intersecting on a string, leading to a
non-chiral effective theory on the string with eight real supercharges
[\papad].  One of our  claims is that the Shiraishi metric in this case is an
example of what is called Hyper-K\"ahler Geometry with Torsion (HKT), which
arises in the context of (4,4) supersymmetric two-dimensional sigma models. 
At this stage, the significance of the torsion is not completely clear, but as
we shall see, it appears naturally in the geometry of the target space of both
one- and two-dimensional supersymmetric sigma models. The
$a=1$, $n=8$ solution lifted to ten dimensions becomes the solution of IIA
supergravity having the interpretation of a wave on a string. This leads to a
chiral effective theory on the string, again with eight real supercharges. Our
claim is that the Shiraishi metric in this case is an  example of what we
shall call Octonionic K\"ahler Geometry  with Torsion (OKT).  This geometry
arises naturally in the context of one-dimensional supersymmetric sigma models
but not in the context of two-dimensional ones.

\chapter{The Supersymmetric Spinning Particle Revisited}

The effective action of black holes that preserve a proportion of
spacetime supersymmetry is described by that of a supersymmetric
spinning particle propagating in a curved background. The background
is determined by the geometry of the moduli space of black holes. Such
an action, up to terms quadratic in velocities, is that of a
one-dimensional supersymmetric sigma model with the black hole moduli
space as target manifold.

It is well known that there is an interplay between the number of
supersymmetries of a supersymmetric sigma model and the geometry of
its target space. Therefore, knowing the amount of supersymmetry
preserved by certain solutions of a supergravity theory, it is
possible to impose strong restrictions on the geometry of their moduli
space. For supersymmetric sigma models in one dimension, extended
supersymmetry imposes weaker conditions on the geometry of the target
space than the same amount of supersymmetry in dimensions two or
higher. This is mainly due to the fact that more couplings amongst the
fields are possible in one dimension, which in higher dimensions are
ruled out by the world-volume Lorentz invariance. Therefore new
geometries can arise on the target space of one-dimensional
supersymmetric sigma models which do not have a direct analogue in
supersymmetric sigma models with world-volume dimensions more than
one. Since we are mainly concerned with the applications of sigma
models to black hole moduli spaces, we shall describe the relation
between the number of supersymmetries in one-dimensional sigma models
and the geometry of their target spaces. For this, we shall begin with
a summary of some of the results of [\coles ] on one-dimensional N=1
supersymmetric sigma models. Then we shall describe the models with
extended supersymmetry generalizing some of the conditions found in [\coles].  For the
relation between the number of supersymmetries in two-dimensional
sigma models and the geometry of their target spaces see
[\gates,\howepap].

The supersymmetry algebra in one dimension is
$$
\{Q^I, Q^J\}= 2 \delta^{IJ} H
\eqn\bone
$$
where $\{Q^I; I=1,\dots,N\}$ are the supersymmetry charges and $H$ is the
Hamiltonian.  In the following we shall describe the cases $N=1, 2, 4$ and $8$.

\section{{\bf  N=1 one-dimensional supersymmetry}}

The simplest case is that of N=1 supersymmetric sigma models, which
have one real supercharge $Q$.  There are several realisations of N=1
supersymmetry in one dimension. For the purpose of this paper it will
suffice to consider a special case of [\coles ] that consists of a
multiplet with a real scalar $X$ and its real fermionic partner
$\lambda$. This is because this sector of the theory determines the
geometry (metric and complex structures) of the moduli space\foot{
Note that for the complete description of the effective theory of a
black hole solution the other sectors in the action of [\coles ] may
have to be included.}.  To describe the geometry associated with the
multiplet $(X,\lambda)$, let the triplet $\big({\cal M}, g, c\big)$ be
a Riemannian manifold ${\cal M}$ with metric $g$ and a 3-form $c$.
The action of such a model is
$$
I={1\over2}\int\, dt\, \big(g_{ij}{d\over dt}X^i\, {d\over dt}X^j+i g_{ij}
\lambda^i\nabla^{(+)}_t\lambda^j
 -{1\over 3!}\partial_{[i}c_{jkl]} \lambda^i\lambda^j\lambda^k\lambda^l\big)
\eqn\btwo
$$
where $X$ are the sigma model fields which are maps from the worldline
to
 a manifold ${\cal M}$ and
 $\lambda$ are worldline real one component fermions which are
sections
 of the bundle $X^*{\cal
TM}\otimes S$ ($S$ is the spin bundle over the worldline). The
covariant
 derivative
$\nabla^{(+)}_t$ is the pull back of the target space covariant derivative
$$
\nabla^{(+)}=\nabla+{1\over2} c
\eqn\bthree
$$
with respect to the map $X$, where  
$$
\Gamma^{(+)}{}^i{}_{jk}=\Gamma^i_{jk}+{1\over2} c^i{}_{jk}\ ,
\eqn\bfour
$$
$\Gamma$ is the Levi-Civita connection of the metric $g$ and
the first index of $c$ is raised with the metric $g$.  Therefore,
 $\nabla^{(+)}$ is a metric
connection with torsion $c$.  This is reminiscent of the situation
that
 arises in (1,0)
supersymmetric two-dimensional sigma models but there is an {\it important
difference:} the torsion here is {\it not necessarily a closed 3-form}.  In
fact it turns out that if $c$ is closed then the action above can be obtained
by reducing the action of (1,0) supersymmetric two-dimensional sigma models.

For reasons that will become apparent later, it is convenient to give an
alternative, but equivalent, description of one-dimensional N=1 sigma models
in terms of superfields.  For this, we introduce a real superfield $X$ which is
a map from the $(1|1)$-dimensional real superspace
$\Xi^{(1|1)}$ with coordinates $\{t; \theta\}$ into
${\cal M}$ with components
$$
X=X|\qquad \lambda=DX|\ ,
\eqn\bfive
$$
where  the vertical line denotes the evaluation of the
associated expression at $\theta=0$ and $D$ is the supersymmetry
derivative,
 {\it i.e.}\
$$
D^2=i{d\over dt}\ .
\eqn\bsix
$$
The action \btwo\ can now be rewritten in terms of the superfield $X$  as
$$
I=-{1\over2}\int\, dt d\theta\, \big(ig_{ij}DX^i {d\over dt}X^j+
 {1\over 3!} c_{ijk} DX^i DX^j
DX^k\big)\ . 
\eqn\bseven
$$
It is clear that this action is manifestly N=1 supersymmetric.

\section{{\bf  N=2 one-dimensional supersymmetry}}

Next, let us consider N=2 supersymmetric one-dimensional sigma models.  As in
the case of N=1 supersymmetry, there are many realizations of N=2 supersymmetry
in one dimension [\coles ]. However here we shall describe two special cases.
To distinguish between them,  we shall call the first one 
$N=2a$ and the second one
$N=2b$.  To describe the first realisation, we introduce a real superfield
$X$ which is a map from the real
superspace $\Xi^{(1|2)}$ with coordinates $\{t; \theta^1, \theta^2\}$
into the manifold ${\cal M}$.  The components of this superfield are
$$
X=X|\qquad \lambda=D_1X|\qquad \psi=D_2X|\qquad F=D_1D_2X|\ ,
\eqn\missing
$$
where $\lambda, \psi$ are the fermionic partners of the boson $X$, $F$
 is an auxiliary field and
$D_1, D_2$ are the supersymmetry derivatives, {\it i.e.}\
$$
D_1^2=i{d\over dt}\ ,
\qquad
D_2^2=i{d\over dt}\ ,
\qquad
D_1D_2+D_2D_1=0\ .
\eqn\beight
$$
The most general action of the $N=2a$ multiplet is
$$
I={1\over 2}\int dt\, d^2\theta\, \big((g+b)_{ij} D_1X^i
D_2X^j+\ell_{ij}
D_1X^i D_1X^j+m_{ij}D_2X^i
D_2X^j\big)\ ,
\eqn\bnine
$$
where $b_{ij}, \ell_{ij}, m_{ij}$ are two-forms on ${\cal M}$. If the
couplings $\ell, m$ vanish, then this action is the reduction of the usual
two-dimensional (1,1)-supersymmetric sigma model; the couplings
$\ell, m$ correspond to non-Lorentz invariant terms in the two-dimensional
action. In particular, the torsion in this case is a ${\it closed}$ three-form
of the sigma model manifold. In what follows we shall assume that the
couplings $\ell, m$ vanish since they do not enter in the applications to the
black hole moduli spaces.

The $N=2b$ one-dimensional supersymmetry is associated with a {\it
complex chiral} superfield $Z$. These are most easily described by
starting with a real superfield as above and by imposing the condition
$$
D_2X^i= I^i{}_j D_1X^j\ ,
\eqn\bten
$$
where $I$ is an endomorphism of the tangent bundle of ${\cal M}$.
Consistency of this constraint with the differential algebra of the
 supersymmetry operators
\beight\ implies that 
$$
\eqalign{
I^2&=-1
\cr
N(I)&=0\, }
\eqn\beleven
$$
where $N(I)$ is the Nijenhuis tensor of $I$.  Both these conditions imply
that the endomorphism $I$ is an  (integrable) {\sl complex structure}. 
Adopting complex coordinates
 on ${\cal M}$ with respect to $I$, we can write the above constraint as
$$
\bar \Delta Z=0\ ,
\eqn\btwelve
$$
where $X=(Z, \bar Z)$ in complex coordinates and $\Delta=D_2+iD_1$. 
 The components of $Z$ are 
$$
Z=Z| \qquad \lambda=\Delta Z|\ ,
\eqn\bthirteen
$$ 
where $Z$ is a complex boson and $\lambda$ is a complex fermion. This
multiplet is associated with the two-dimensional (2,0) multiplet. We
remark that the corresponding sigma model target space of the N=2b
multiplet is a complex manifold, unlike the sigma model target space of
the N=2a multiplet described above which is real one.  To determine
the conditions on the couplings of the action \bseven\ required by N=2
supersymmetry, we follow [\howepap] and express the second
supersymmetry transformation in terms of the N=1 superfield $X$ as
$$
\delta X^i=\eta\, I^i{}_j DX^j\ ,
\eqn\bfourteen
$$ 
where $\eta$ is the parameter of the transformation.
A straightforward computation reveals that the action \bseven\ is
invariant
 under this
transformation provided\foot{For another derivation of the invariance of
the action conditions see [\sfetsos].} that
$$
\eqalign{
g_{k\ell} I^k{}_i I^\ell{}_j&=g_{ij}
\cr
\nabla^{(+)}_{(i} I^k{}_{j)}&=0
\cr
\partial_{[i}\big(I^m{}_j c_{|m|kl]}\big)-
2 I^m{}_{[i} \partial_{[m} c_{jkl]]}&=0\ .}
\eqn\bfifteen
$$
The first condition is the usual hermiticity condition of the metric
 $g$ with respect to the
complex structure $I$.  An alternative way to write the second condition is
$$
\nabla^{(+)}_{i} I_{jk}=\nabla^{(+)}_{[i} I_{jk]}\ ,
\eqn\bsixteen
$$
lowering the index of the complex structure with the metric $g$. If
the torsion $c$ vanishes, then this condition becomes the Yano tensor
condition which has already appeared in the context of one-dimensional
supersymmetric sigma models in [\gibvan]. Therefore, the condition
\bsixteen\ is a generalisation of the Yano tensor condition for a
connection with torsion.  The last condition in
\bfifteen\ does not have a direct geometrical interpretation. 
 For convenience, we shall use the
form notation\foot{Our normalization convention for a p-form,
$\omega$, is $\omega={1\over p!} \omega_{i_1\dots i_p} dx^{i_1}
\wedge\dots \wedge dx^{i_p}$.}  
$$
\iota_Idc-{2\over3} d\iota_Ic=0
\eqn\missinga
$$
for this relation, where $\iota_I$ is the inner derivation with
respect to the the complex structure $I$.

It is instructive to compare the conditions that we have found on the
geometry of the target space of one-dimensional sigma models with N=2b
supersymmetry with those of two-dimensional sigma models with (2,0)
supersymmetry. The first condition in \bfifteen\ also arises in the
context of two-dimensional (2,0)-supersymmetric sigma models.  The
last two conditions in \bfifteen\ do not have a direct two-dimensional
interpretation.  In fact in two dimensions, both conditions are replaced by the
covariant constancy condition of $I$,
$$
\nabla^{(+)}_i I^j{}_k=0\ ,
\eqn\bseventeen
$$
with respect to the $\nabla^{(+)}$ connection.  It turns out this is
a much stronger condition and any solution of the covariant constancy
condition solves the last two conditions in \bfifteen.  However the
converse is not true. Finally, we remark that it is straightforward to
write an off-shell superfield action for the N=2b multiplet using the
method of [\howepap, \coles] but we shall not present it here.

\section{{\bf  N=4 one-dimensional supersymmetry}}

Next, let us turn to investigate the one-dimensional N=4
supersymmetric sigma models. We shall again describe two special
multiplets that we shall call N=4a and N=4b, respectively.  The former
is the reduction of the two-dimensional (2,2) supersymmetry multiplet
and the latter is associated with the two-dimensional (4,0)
supersymmetry multiplet.  Since the N=4a multiplet is the reduction of
the two-dimensional (2,2) one, the geometry of the target manifold of
the one-dimensional sigma model in this case is the same as that of
the two-dimensional one; we shall not repeat the analysis here (see
for example [\howepap]).  The geometry associated with the N=4b
multiplet is not necessarily the hyper-K\"ahler with torsion (HKT)
geometry of the two-dimensional (4,0) multiplet. To find the
conditions on the geometry of the target space required by the N=4b
multiplet, we use N=1 superfields to write the extended supersymmetry
transformations as
$$
\delta X^i=\eta^r I_r{}^i{}_j DX^j
\eqn\abone
$$
where $\{\eta^r; r=1,2,3\}$ are the supersymmetry parameters and
$\{I_r; r=1,2,3\}$ are
endomorphisms of the tangent bundle of the sigma model manifold.
The conditions from the closure of the N=4 supersymmetry algebra are
$$
\eqalign{
I_r I_s+I_s I_r&=-2 \delta_{rs}
\cr
N(I_r, I_s)&=0}
\eqn\abtwo
$$
and the conditions from the invariance of the action are
$$
\eqalign{
g_{k\ell} I_r{}^k{}_i I_r{}^\ell{}_j&=g_{ij}
\cr
\nabla^{(+)}_{(i} I_r{}^k{}_{j)}&=0
\cr
\iota_rdc-{2\over3}d\iota_rc&=0\ ,}
\eqn\abthree
$$
where $N(I_r, I_s)$ is the Nijenhuis tensor for the pair of
endomorphisms $(I_r, I_s)$ (see for example [\howepapc]) and $\iota_r$
denotes inner derivation with respect to the endomorphism $I_r$.
Therefore, the target manifold admits three complex structures which
have vanishing mixed Nijenhuis tensors and obey the algebra of a basis
in ${\rm Cliff} (\Bbb E^3)$ equipped with a negative definite inner
product.  In fact, the three complex structures in this case obey the
algebra of imaginary unit quaternions.  This is because if one is
given two anticommuting complex structures $I_1, I_2$, one can
construct a third one $I_3$ by multiplying the two together, {\it
i.e.}\ $I_3=I_1 I_2$.  In addition, the metric is hermitian with
respect to all complex structures.  The last two conditions in
\abthree\ are the analogues of the last two conditions
\bfifteen\ but now for each complex structure.

It is instructive to compare these conditions with those of HKT
manifolds
 [\howepapd].   A {\it
weak} HKT manifold is a Riemannian manifold
$\{{\cal M}, g , c\}$ equipped with a metric $g$, a three-form $c$ and three
complex structures
$\{I_r; r=1,2,3\}$ that obey the following compatibility conditions: 
\item {(i)} The complex structures
obey the algebra of imaginary unit quaternions
$$
I_r I_s=- \delta_{rs}+\epsilon_{rst} I_t\ ,
\eqn\abfour
$$

\item{(ii)} the metric is hermitian with respect to all complex structures
$$
g_{k\ell}\, I_r{}^k{}_i I_r{}^\ell{}_j=g_{ij}
\eqn\abfour
$$
(no summation over the index $r$) and 

\item{(iii)} the complex structures are covariantly constant with
	respect to the
$\nabla^{(+)}$ covariant derivative
$$
\nabla^{(+)}_k I_r{}^i{}_j=0\ .
\eqn\abfive
$$
\noindent If in addition the three-form $c$ is closed, then ${\cal M}$ has a
{\it strong} HKT structure.  In the classical theory, the target space of
two-dimensional (4,0)-supersymmetric sigma models has a strong HKT
structure\footnote\dag{It was observed in [\howepapd], though, that in the
quantum theory the target space
becomes a weak HKT manifold due to the anomaly cancellation mechanism.}.

The main difference between HKT manifolds and those arising in the
context of one-dimensional N=4b supersymmetric sigma models is that
the covariant constancy condition of the complex structures in the
weak HKT geometry is replaced by the last two conditions in \abthree.
It turns out that the covariant constancy condition implies those of
\abthree.  Therefore {\it any weak} HKT manifold {\it solves} all the
conditions required by N=4b one-dimensional supersymmetry.

We can write a one-dimensional N=4b supersymmetry multiplet in superspace as
$$
D_rX^i=I_r{}^i{}_j D_0X^j
\eqn\absix
$$
where $X^i$ are maps from the superspace $\Xi^{(1|4)}$ with
coordinates
 $\{t, \theta^0, \theta^r;
r=1,2,3\}$ into the sigma model target space ${\cal M}$ and
 $\{D_0, D_r; r=1,2,3\}$ are
the supersymmetry derivatives obeying the algebra
$$
\eqalign{
D_0^2&=i{d\over dt}
\cr
D_0D_r+D_r D_0&=0
\cr
D_sD_r+D_r D_s&=2i\delta_{rs}{d\over dt}\ .}
\eqn\abseven
$$             
An action for this multiplet is 
$$
I=-{1\over2}\int\, dt d\theta^0\, \big(i g_{ij}D_0X^i {d\over dt}X^j+
 {1\over 3!} c_{ijk} D_0X^i
D_0X^j D_0X^k\big)\ . 
\eqn\abeight
$$
Note that although this action is not a full superspace integral, 
it is N=4b supersymmetric provided
that the couplings satisfy the conditions \abtwo\ and \abthree.

\section{{\bf  N=8 one-dimensional supersymmetry}}

Finally, let us consider the one-dimensional sigma models with N=8
supersymmetry.  Again we shall describe two special N=8 multiplets
which we shall call N=8a and N=8b.  The former multiplet is the
reduction of the two-dimensional (4,4) supersymmetry multiplet and the
latter is associated with the two-dimensional (8,0) supersymmetry
multiplet.  Since the N=8a multiplet is the reduction of the
two-dimensional (4,4) one, the geometry of the target manifold of the
one-dimensional sigma model in this case is the same as that of the
two-dimensional one; we shall not repeat the full analysis here (see
for example [\howepap]). In particular, if the supersymmetry algebra
closes off-shell, then the sigma model target space ${\cal M}$ admits
two {\it commuting strong} HKT structures.  One of the HKT structures
is with respect to the connection
$$
\nabla^{(+)}=\nabla+{1\over2}c
\eqn\cone
$$ 
and the other is with respect to the connection
$$
\nabla^{(-)}=\nabla-{1\over2}c\ .
\eqn\ctwo
$$
 An action for the N=8a multiplet can be written in a way similar
to that for the (4,4) multiplet in two dimensions [\howepap ].   

Next, let us turn to examine the N=8b case. To find the
conditions required by N=8b supersymmetry,  we use N=1 superfields to
 write the extended
supersymmetry transformations as
$$
\delta X^i =\eta^a I_a{}^i{}_j DX^j
\eqn\cthree
$$
where $\{ \eta^a; a=1,\dots, 7\}$ are the supersymmetry parameters.  
The conditions required by
the closure of the supersymmetry algebra are
$$
\eqalign{
I_a I_b+I_b I_a&=-2 \delta_{ab}
\cr
N(I_a, I_b)&=0}
\eqn\cfour
$$
and the conditions required by the invariance of the action are
$$
\eqalign{
g_{k\ell} I_a{}^k{}_i I_a{}^\ell{}_j&=g_{ij}
\cr
\nabla^{(+)}_{(i} I_a{}^k{}_{j)}&=0
\cr
\iota_adc-{2\over3}d\iota_ac&=0\ ,}
\eqn\cfive
$$
where $\iota_a$ denotes inner derivation with respect to the
endomorphism $I_a$.  The endomorphisms $\{I_a\}$ are complex
structures that obey the algebra of a basis in ${\rm Cliff}(\Bbb E^7)$
equipped with a negative definite inner product; the underlying
algebraic structure is naturally associated with that of
octonions. The remaining conditions are similar to those of the N=4b
multiplet but this time they apply to seven complex structures instead
of three.  We shall call a Riemannian manifold $\{{\cal M}, g, c\}$
equipped with metric $g$, antisymmetric tensor $c$, and complex
structures $\{I_a\}$ that obey the compatibility conditions \cfour\
and \cfive\ an {\it Octonionic K\"ahler with Torsion} manifold, or OKT
for short.

This appears to be the appropriate generalisation of the HKT structure
in the (8,0)-supersymmetric
context.  To see this, observe that in the HKT structure, the last two
conditions in \cfive\ are replaced with the  covariant constancy condition 
$$
\nabla_i^{(+)}(I_a)^j{}_k=0\ ,
\eqn\csix
$$
of the complex structures. Now if the manifold is eight dimensional, then the
curvature of $\nabla^{(+)}$ must vanish because the complex structures form an
irreducible representation of ${\rm Cliff}(\Bbb E^7)$. This is very
restrictive and one of the reasons behind  the absence of known examples of
interacting two-dimensional (8,0)-supersymmetric  sigma models.  However as we
shall see, there are examples of non-trivial OKT manifolds, which moreover
serve as moduli spaces of n=8 a=1 black holes preserving
$1/4$ of the supersymmetry of IIA supergravity.

To describe an off-shell N=8b superspace multiplet, let $X$ be a map
from the $\Xi^{(1|8)}$ superspace with coordinates $\{t; \theta^0,
\theta^a, a=1,\dots,7\}$ into an OKT manifold ${\cal M}$.  Then we
impose the constraints
$$
D_aX^i=I_a{}^i{}_j D_0X^j
\eqn\cseven
$$
where $\{D_0, D_a; a=1,\dots,7\}$ are the supersymmetry derivatives
along the corresponding
Grassmann directions in $\Xi^{(1|8)}$. The algebra of $\{D_0, D_a;
a=1,\dots,7\}$ is a direct generalization of that in   \abseven\ and we shall
not present it here.  An action for this multiplet is 
$$
I=-{1\over2}\int\, dt d\theta^0\, \big(i g_{ij}D_0X^i {d\over dt}X^j+ 
{1\over 3!} c_{ijk} D_0X^i
D_0X^j D_0X^k\big)\ . 
\eqn\ceight
$$
Note that although this action is not a full superspace integral, it
is N=8b supersymmetric provided that the couplings in the action
satisfy the conditions \cfour\ and \cfive.  We remark that this action
is similar to that of the N=4b multiplet above but that the coupling
constants in the action of the N=8b multiplet obey different
conditions from those of the N=4b case.

%%%%%%%%%%%%%%%%%%%%%%%%%%%%%%%%%%%%%%%%%%%%%%%%%%%%

\chapter{ Hyper-K\"ahler Geometry with Torsion}

A {\it strong} Hyper-K\"ahler Geometry with Torsion $\{ {\cal M}, g, c \} $
may be described in a number of ways. An equivalent but perhaps more concise
description than the one of the previous section is to say that it consists of 
\medskip \item {(i)} a $4k$-dimensional hypercomplex manifold ${\cal M}$, a
compatible metric $g$ with associated Levi-Civita connection $\nabla$ and
\medskip \item {(ii)} a closed 3-form $c$ such that if we use the inverse
metric to convert $c$ to a vector valued 2-form, which we also call $c$,
then the metric preserving affine connection:
$$
\nabla ^{(+)} =\nabla + {1\over2} c
\eqn\done
$$
preserves the hypercomplex structure.

We now expand a little on this definition. Firstly a hypercomplex
manifold is one admitting three integrable complex structures, $\{I_r;
r=1,2,3\}=\{I, J, K\}$ satisfying the algebra of the imaginary
quaternions. Alternatively one may say that the structural group of
the tangent bundle $T{\cal M}$ may be reduced from $GL(4k; {\Bbb R})$
to $GL(k; {\Bbb H})$. A compatible metric is one for which $I,J,K$ are
isometries. In other words $g$ is Hermitian with respect to all three
complex structures. Given $g$ we may construct three 2-forms $(\omega
_I, \omega _J, \omega _K)$ from $I,J,K$ by index lowering.  If we were
dealing with a Hyper-K\"ahler structure, all three 2-forms would be
closed and the holonomy of the Levi-Civita connection $\nabla$ would
lie in $Sp(k) \subset SO(4k; {\Bbb R})$. For a Hyper-K\"ahler Geometry
with torsion we demand something weaker: merely that the holonomy of
the metric preserving affine connection $\nabla ^{(+)} =\nabla +
{1\over2} c $ lies in $Sp(k) \subset SO(4k; {\Bbb R})$. Equivalently
one demands that
$$
d\omega _I -\iota_I c=0\ ,\qquad  d\omega _J -\iota_J c=0\ , 
\qquad d\omega _K -\iota_K c=0\ .
\eqn\dtwo
$$

Evidently one may construct products of HKT structures so as to obtain
HKT structures on the product. One also has a natural notion of a
group $G$ of symmetries of an HKT structure. One may also restrict an
HKT structure to a totally geodesic hyper-complex submanifold $\Sigma$
of an HKT manifold. This requires that all vectors tangent to $\Sigma$
remain tangent to $\Sigma$ when acted upon by $I,J$ and $K$ One also
requires that the connection $\nabla^{(+)}$ is used to propagate
vectors initially parallel to $\Sigma$ such that they remain parallel
to $\Sigma$. A convenient way to identify a totally geodesic
hyper-complex submanifold is as the fixed point set of a group $G$ of
symmetries. In terms of sigma models, totally geodesic submanifolds
arise by imposing constraints on the sigma model which commute with
the action of supersymmetry and allow its consistent truncation to a
model with fewer fields.

There are two basic examples of HKT structures, the flat structure
with zero torsion on ${\Bbb H}$ and the Wess-Zumino-Witten model on
${\Bbb H}\\ \{0\} \equiv {\Bbb H} ^\star$. Both admit $(4,4)$
supersymmetry (for the latter case see [\rocek]). In the natural
quaternionic notation, adapted to one hyper-complex structure, the
metric in the first case may be expressed as
$$
ds^2 = dq d{\bar q}
\eqn\dthree
$$
with vanishing torsion and in the second
by
$$
ds^2 =  { dq d{\bar q} \over q {\bar q} }.
\eqn\dfour
$$
while the torsion three-form corresponds to the volume form on the
unit three sphere. To pass to the other hypercomplex structure, one
takes the quaternionic conjugate. Topologically the Wess-Zumino-Witten
model is defined on ${\Bbb R} \times S^3$, the universal covering
space of what mathematicians call the Hopf surface $S^1 \times S^3$.
This is a well known example of a complex manifold which does not
admit a K\"ahler structure.  The metric is the product metric on
${\Bbb R} \times S^3$ and the 3-from $c$ is the volume form on
$S^3$. If one identifies $S^3$ with $SU(2)$ and defines $2t = \log (q
{\bar q})$, then the connection $\nabla ^{(+)}$ is given by
$$
{\partial \over \partial t} +\nabla^{(+)}_{SU(2)}
\eqn\dfive
$$
where  $ \nabla^{(+)} _{SU(2)}$ is the standard  connection
on the Lie group $SU(2)$ defined using, say, the right translations.
The three two forms obtained from the complex structures by index lowering are
$$
\omega _r = dt \wedge \sigma_r + d \sigma _r\ ,
\eqn\dsix
$$
where $r=1,2,3$ and $\sigma_r$ are left-invariant one-forms on
$SU(2)$,
{\it i.e.}\
$d\sigma_r={1\over2} \epsilon^r{}_{st} \sigma^s\wedge \sigma^t$.

The Wess-Zumino-Witten model admits the orientation preserving
symmetry 
$$
R:\quad q \rightarrow -q
\eqn\dseven
$$
preserving both  $HKT$ structures. In effect of $R$ is the antipodal
map on the 3-sphere factor which is orientation preserving. 
It therefore leaves the volume form and hence the torsion invariant.

\section{{\bf Multi-Models}}

In order to obtain more complicated models, including the Shiraishi 
metric for the relative moduli of the n=4, a=1 black holes, we
take products of the above two basic
models:
$$\bigl ( {\Bbb H}\bigr )  ^v \times \bigl ({\Bbb H}^\star\bigr )^e
\eqn\deight
$$
with coordinates $w_a$ and $q_i$ respectively, with metric
$$
ds^2 = \sum ^v dw_a d {\bar w}_a + \sum ^e { d q_i d {\bar q}_i
 \over q_i {\bar q}_i}
\eqn\dnine
$$
and we impose some constraints. Consider for example the case $v=e=1$.
Dropping the indices on $w$ and $q$ we impose
$$
w-q=0
\eqn\dten
$$
and recognize the well known  metric on ${\Bbb H}^\star$, the 
transverse space of a single solitonic  5-brane  [\calhar] . 
This corresponds to the
Shiraishi metric on the relative  moduli space of two $a=1, n=1\,$ 0-branes,
or by lifting to six dimensions to obtain two self-dual strings.  
Note that the constraint restricts us to the fixed point set
of the ${\Bbb Z}_2$ action:
$$
(w+q,w-q) \rightarrow (w+q, -w+q).
\eqn\deleven
$$
Geometrically the ${\Bbb Z}_2$ action is reflection in the hyperplane
defined by the constraint.

To get the transverse metric of the $k$-5-brane 4-metric 
one takes $v=1$ and $e=k$ and imposes the $k$ constraints 
$$
w+a_i-q_i=0\ ,
\eqn\dtwelve
$$
where the $a_i$ are $k$ constant quaternions. These complete
asymptotically Euclidean 4-metrics on ${\Bbb H} \\ \cup_i\{a_i\}$ were
proposed as gravitational instantons by D'Aurilia and Regge [\aurreg]
and dubbed axionic instantons by Rey [\rey].

To get the Shiraishi metrics we take $v=k$ and $e=  {1\over 2}k(k-1)$
and replace the index $i$ on $q_i$ by the compound index 
$ab$ with $0<a<b\le k$. The constraints are
$$
w_a-w_b- q_{ab}=0.
\eqn\dthirteen
$$
To be more precise, agreement with the Shiraishi metrics requires an
appropriate rescaling of the coordinates so as to introduce the
various \lq\lq mass parameters" that appear in the black hole moduli
metrics.  This completes our demonstration that the moduli space of
self-dual strings in six dimensions admits $(4,4)$ supersymmetry.

More general metrics will be studied later. Before doing so we want to
comment on the similarity between the construction just given for HKT
geometries associated to the Shiraishi metrics with $a=1$ in $n+1=5$
spacetime dimensions and the Hitchin -Karlhede-Lindstrom-Rocek
quotient construction of the Hyper-K\"ahler geometries associated to
the Shiraishi metrics with $a=1$ in $n+1=4$ spacetime dimensions
[\hitchin]. The moment map constraints used in the latter construction
involve 3-vectors instead of 4-vectors, but are otherwise
identical. It seems that this is just a particular case of
T-duality. The 4-dimensional metrics lift to 6-branes and the
5-dimensional metrics lift to five branes in 10 spacetime
dimensions. From the point of view of the moduli space the relative
moduli space in the case of two of 6-branes is the 3-metric obtained
by taking the ordinary quotient of the Taub-NUT metric by a
triholomorphic $U(1)$ and in the case of the 5-branes it is the
axionic instanton 4-metric.
 
Note the amusing fact that taking the moduli space of the 
axionic instantons, regarded as Neveu-Schwartz 5-branes, leads to
a flat metric.

\section {{\bf Sigma Model Duality}}

An alternative method of constructing HKT geometries is to use the
close relationship between hyper-K\"ahler and HKT geometries.  This is
most easily demonstrated in four dimensions where starting from a
hyper-K\"ahler geometry with metric $ds^2_{hk}$, one can obtain a weak
HKT one using the Callan-Harvey-Strominger ansatz\footnote\dag{Not all
4-dimensional HKT geometries can constructed in this way [\bonneau,
\papadb].}[\calhar]
$$
\eqalign{
ds^2&=H ds^2_{hk}
\cr
c&=3\, *dH\ ,}
\eqn\daone
$$
where $H$ is a function on the hyper-K\"ahler manifold and the Hodge
 duality operation has been taken with respect to the hyper-K\"ahler
metric.  The Callan-Harvey-Strominger ansatz gives a {\sl strong} HKT
geometry provided that $H$ is a harmonic function with respect to the
hyper-K\"ahler metric, {\it i.e.}\
$$
\nabla^2_{hk}H=0\ .
\eqn\datwo
$$
The complex structures of the HKT geometry are those of the
hyper-K\"ahler geometry.  For example, if we choose as a
hyper-K\"ahler manifold ${\Bbb E}^4$ then the associated HKT geometry,
with
$$
H=1+\sum_i {\mu_i\over 2 |x-x_i|^2}\ ,
\eqn\dathree
$$
is that of the solitonic five-brane or that of the relative moduli
space of two $n=4$, $a=1$ black holes described in the previous
sections.  The HKT metric is complete and asymptotically flat. At the
centres of the harmonic function there are infinite throats isometric
to ${\Bbb R} \times S^3$.  It is worth pointing out that the
associated one- or two-dimensional sigma models may have N=8a or (4,4)
off-shell supersymmetry, respectively.  This is because in ${\Bbb E
}^4$, we can introduce two commuting triplets of complex structures
with each triplet obeying the algebra of imaginary unit
quaternions. The associated K\"ahler 2-forms of the first triplet is a
basis of the self-dual two forms in ${\Bbb E}^4$ and, similarly, the
associated K\"ahler 2-forms of the second triplet is a basis of the
anti-self-dual two forms in ${\Bbb E}^4$.

This relation between hyper-K\"ahler and HKT geometries can be
extended beyond four dimensions using sigma-model duality [\buscher].
Sigma-model duality is an operation which is applied to the metric,
torsion and dilaton couplings of a two-dimensional sigma model
admitting an $U(1)$ isometry.  The effect of the operation is to give
another quantum-mechanically equivalent sigma model with a $U(1)$
isometry but with couplings different from those of the original
model.  For what follows, it is sufficient to investigate the effect
of sigma-model duality on a bosonic sigma model with just a metric
coupling.  For this, let us suppose that the sigma model metric
$$
ds^2=V^{-1} (d\tau +\omega)^2+V \gamma_{ij} dx^i dx^j
\eqn\dafour
$$
admits a Killing vector $X={\partial/\partial \tau}$, where $\gamma$ is the
metric on the space of orbits of the isometry.  Performing sigma model
duality  along $X$, we find that the couplings of the dual model are
$$
\eqalign{
ds^2&= V (d^2\tau+ \gamma_{ij} dx^i dx^j)
\cr
c&=-3\, d\tau\wedge d\omega
\cr
e^{2\Phi}&= V\ ,
 }
\eqn\dafive
$$
where $\Phi$ is the `dilaton'.  If the dimension of the target space
of the sigma model is less than or equal to nine, then sigma model
duality coincides with the T-duality of the common
Neveu-Schwartz$\otimes$Neveu-Schwartz sector of the various string
theories.  The main point to observe is that, although the original
sigma model has just a metric coupling, after T-duality we find that
the dual model has, apart from the metric coupling, a non-zero
Wess-Zumino term and a `dilaton'.  T-duality under certain conditions
preserves supersymmetry. Since two-dimensional sigma models with
hyper-K\"ahler metrics admit (4,4) supersymmetry, the dual models also
admit (4,4) supersymmetry, and hence their target space has two copies
of a {\it strong} HKT structure. In what follows, we shall neglect the
`dilaton' because it is not necessary for determining the geometry of
the moduli space of black holes.

As an example, let us find the HKT structure associated to the
Gibbons-Hawking hyper-K\"ahler metric [\gibhawk, \gibruba]
$$
ds^2= H^{-1} (d\tau +\omega)^2+H ds^2(\bE^3)\ ,
\eqn\dasix
$$
where
$$
\star dH=- d\omega\ ,
\eqn\daseven
$$
{\sl i.e.}\ $H$ is a harmonic function on three-dimensional 
 Euclidean space ${\Bbb E}^3$.
This metric admits a tri-holomorphic Killing vector field 
$X={\partial/\partial \tau}$. After
an appropriate identification of the $\tau$ coordinate, 
the metric is geodesically complete.
Applying T-duality to this metric leads to
$$
\eqalign{
ds^2&= H ds^2({\Bbb E}^4)
\cr
c&=3\,*dH\ ,
}
\eqn\daeight
$$
where the Hodge duality operation has been taken with respect to the
flat metric on $\Bbb E^4$, (see also [\bakas, \lust]). This HKT
geometry is similar to that derived above using the ansatz \daone\ and
with the four-dimensional Euclidean space ${\Bbb E} ^4$ as a starting
hyper-K\"ahler manifold. However, there is a difference in that the
conformal factor $H$ in \daeight\ is a harmonic function on ${\bE}^3$
rather than a harmonic function on ${\bE}^4$ which is the case in
\daone. A consequence of this is that the HKT geometry \daeight\ is
incomplete.  However, it is clear that we can find a complete geometry
associated to \daeight\ by allowing the harmonic function $H$ to be
harmonic on ${\bE}^4$.

%%%%%%%%%%%%%%%%%%%%%%%%%%%%%%%%%%%%%%%%%%%%%%%%%%%%%%%%%%%%%%%%%%%

\section{{\bf HKT geometry in various dimensions}}

To find more general HKT geometries in $4k$ dimensions, $k>1$, from
those in section (4.1), we shall begin with a class of hyper-K\"ahler
geometries that admit a $U(1)^k$ group of triholomorphic
isometries. To describe these metrics, let us decompose the $4k$
coordinates $\{y^M; M=1,\dots, 4k\}$ of the hyper-K\"ahler manifold as
$$
y^M=(\tau_i, x^{ri})\ ;\qquad  r=1,2,3\ ,\qquad  i=1,\dots, k\ ,
\eqn\daaaone
$$
where $\tau_i$ are the coordinates adapted to the isometries, {\it
i.e.}\ the Killing vector fields are $X^i=\partial/\partial\tau_i$.
Then the hyper-K\"ahler metric can be written as follows:
$$
ds^2=U^{ij} (d\tau_i+{\bfomega}_{ik}\cdot d{\bf x} ^k ) (d\tau_j+
{\bfomega} _{j\ell}\cdot d{\bf x} ^\ell)+U_{ij} d{\bf x}^i\cdot
d{\bf x} ^j
\eqn\dbone
$$
where $(\cdot)$ denotes the inner product with respect to the flat metric in
${\Bbb E}^3$ and the coefficients $U$ are functions only of $x$;
$U=U(x)$. Moreover, they satisfy 
$$
\eqalign{
U_{ij}&=U_{ji}
\cr
\star dU_{ij}&=-d\bfomega_{ij}\ ,}
\eqn\dbtwo
$$
where the Hodge star operation is taken with respect to the flat metric in
${\Bbb E}^3$.
To find explicit solutions to the above conditions \dbtwo, we follow
[\town] and write
$$
U_{ij}=U^\infty_{ij}+\Delta U_{ij}\ ,
\eqn\dbthree
$$
where $U^\infty$ is the asymptotic value of $U$ as $|{\bf x}^k|$ goes
to infinity.  Then
$\Delta U_{ij}$ is the sum of terms of the form
$$
{p_i p_j\over |p_k {\bf x}^k- {\bf a}|}\ ,
\eqn\dbfour
$$
for different values of $p$ and different centres $a$. Therefore, 
the most general form of $\Delta
U_{ij}$ is
$$
\Delta U_{ij}=\sum_{\{p\}}\sum_{{\bf a}} {p_i p_j\over |p_k 
{\bf x}^k- {\bf a}(\{p\})|}
\eqn\dbfive 
$$
It is well-known that this geometry 
can be characterised by the properties of the co-dimension three planes
$$
p_k {\bf x}^k-{\bf a}=0\ ,
\eqn\dbsix
$$
in ${\bE}^{3k}$.  In particular, the metric is non-singular provided that
$\{p_1, \dots, p_k\}$ are co-prime integers, and the various $(3k-3)$-planes
do not coincide and intersect only pairwise. 

There is a chain of strong HKT structures associated with the above
hyper-K\"ahler geometry. Each HKT structure in the chain can be found
by T-dualizing the hyper-K\"ahler geometry along one or more Killing
vector field directions $\tau_i$.  The case that we shall present here
is the one that arises after dualizing all Killing vector directions
once.  The resulting strong HKT geometry is
$$
\eqalign{
ds^2&=U_{ij} \big(d\tau^i\, d\tau^j+ d{\bf x}^i\cdot d{\bf x}^j\big)
\cr
c&={3\over2}\,\epsilon_{rs}{}^t\, \partial_{ti}U_{jk}\,
 d\tau^i\wedge dx^{rj}\wedge dx^{sk} \ .}
\eqn\dbseven
$$
This metric is singular. However as in the 4-dimensional case, 
we can extend the dependence of the
coefficient $U_{ij}$ of the metric.  For this, we decompose the
$4k$-coordinates $\{y^M; M=1,\dots, 4k\}$ of the HKT manifold as 
$$
y^M=x^{\mu i}\qquad i=1,\dots, k\qquad \mu=0,\dots,3\ .
\eqn\dbeight
$$ 
The new metric and torsion can be written as
$$
\eqalign{
ds^2&=U_{ij}\, d{\bf x}^i\cdot d{\bf x}^j
\cr
c&={1\over2}\, \epsilon_{\mu\nu\lambda}{}^\rho\, 
\partial_{\rho i} U_{jk}\, dx^{\mu i}\wedge
dx^{\nu j}\wedge dx^{\lambda k} \ ,}
\eqn\dbnine
$$
where $U$ is a function of ${\bf x}^k$ and $\epsilon$ is
 the Levi-Civita tensor with respect
to the flat metric on
$\Bbb E^4$. For the torsion $c$ to be a 3-form, we require 
$$
\eqalign{
U_{ij}&=U_{ji}
\cr
\partial_{i\mu} U_{jk}&=\partial_{j\mu} U_{ik}\ , }
\eqn\dbten
$$ 
where Hodge star operation is with  respect to the flat metric
in $\Bbb E^4$.  If we further
require $c$ to be a closed 3-form, then 
$$
\partial_i\cdot \partial_j U_{kl}=0\ .
\eqn\dbeleven
$$

To find explicit examples of strong HKT geometries, we write
$$
U_{ij}=U^\infty_{ij}+\Delta U_{ij}\ ,
\eqn\bdtwelve
$$
where  $U^\infty_{ij}$ is the asymptotic value of $U_{ij}$ as
 $|{\bf x}^k|$ goes to infinity. Then, 
$\Delta U_{ij}$ is a sum of terms of the form
$$
{p_i p_j\over |p_k {\bf x}^k- { \bf a}|^2}={1\over4}\partial_i
\cdot\partial_j \log|p_k {\bf x}^k-
{\bf a}|^2
\eqn\dbthirteen
$$
for different choices for the real numbers $p$ and for the centres
$a$.  
Therefore, the most
general form for $\Delta U_{ij}$ is
$$
\Delta U_{ij}={1\over 4}\sum_{\{p\}}\sum_{{\bf a}} 
\partial_i\cdot\partial_j \log|p_k {\bf x}^k-
{\bf a}(\{p\})|^2 \ .
\eqn\bdfourteen
$$
In direct correspondence with the associated hyper-K\"ahler geometries, 
these HKT geometries are
naturally associated with the co-dimension four planes
$$
p_k {\bf x}^k-{\bf a}=0
\eqn\dbfifteen
$$ 
in ${\Bbb E}^{4k}$.  It appears that these metrics are non-singular on
the complement of these planes in $\Bbb E^{4k}$ provided that the
planes, if they intersect, intersect only pairwise.

It remains to give the complex structures of the above HKT geometry.  These are
$$
{\bf I}_r=1\otimes I_r\ ,
\eqn\dbsixteen
$$
where $\{I_r; r=1,2,3\}$ are the three complex structures in ${\Bbb
E}^4$ associated with, say, a basis in the space of constant self-dual
2-forms in $\Bbb E^4$.  Then it is straightforward to verify that the
covariant constancy condition of the complex structures with respect
to the $\nabla^{(+)}$ covariant derivative and the closure of $c$
imply only the conditions \dbten\ and \dbeleven. Thus we have directly
shown that $({\cal M},g, c)$ of \dbnine\ with coefficients given in
\bdfourteen\ has a strong HKT structure.  In fact there is another HKT
structure on $({\cal M}, g,c)$ associated with the connection
$\nabla^{(-)}$ and with complex structures $$ {\bf J}_r=1\otimes J_r\
, \eqn\dbseventeen $$ where in this case $\{J_r; r=1,2,3\}$ are the
three complex structures in ${\Bbb E}^4$ associated with a basis in
the space of constant anti-self-dual 2-forms in $\Bbb E^4$.  It turns
out that the two HKT structures commute.  Therefore, the corresponding
one- or two-dimensional sigma model whose target space is the strong
HKT manifold $({\cal M}, g, c)$ may admit {\it off-shell} N=8a or
(4,4) supersymmetry, respectively.

\chapter{ Octonionic K\"ahler Geometry with Torsion}

The algebraic structure underlying the OKT geometry is that of the
 octonions, $\Bbb O$.  Let $\{e_0,
e_a; a=1,\dots, 7\}$ be a basis in $\Bbb O$  consisting of the unit
 octonions. In this basis, we choose
$e_0$ to be the identity, so it commutes with all the other elements of
the basis. The rest of the basis elements satisfy
$$
e_a e_b=-\delta_{ab}+ \varphi_{ab}{}^c e_c
\eqn\eone
$$
where $\varphi_{abc}=\varphi_{ab}{}^d \delta_{dc}$ are the structure
constants of the octonions;
$\varphi$ is antisymmetric in all its indices. Next, let  ${}^*\varphi$
 be the Poincar\'e dual  of
$\varphi$,
$$
{}^*\varphi_{abcd}={1\over 3!} \epsilon_{abcd}{}^{pqr} \varphi_{pqr}\ ,
\eqn\etwo
$$
then
$$
\varphi_{abf} \varphi_{cd}{}^f=\delta_{ac} \delta_{bd}-\delta_{ad} \delta_{bc}-
{}^*\varphi_{abcd}\ ,
\eqn\ethree
$$
and
$$
{}^*\varphi^f{}_{abc}   \varphi_f{}^{de} =6\varphi^{[d}{}_{[ab} 
\delta^{e]}_{c]}\ . 
\eqn\efour
$$

We can use the structure constants of the octonions to introduce
 seven complex structures in $\Bbb
E^8$  as follows:
$$
\eqalign{
(I_a){}^0{}_b&= \delta_{ab}
\cr
(I_a){}^b{}_0&= -\delta_a^b
\cr
(I_a){}^b{}_c&=\varphi_a{}^b{}_c\ . }
\eqn\efive
$$
It is straightforward to see (i) that the flat metric in $\Bbb E^8$ is
hermitian with respect to all the complex structures and (ii) that these
complex structures obey the gamma-matrix relations of a basis in ${\rm
Cliff}(\Bbb E^7)$ equipped with a negative-definite inner product.  It is then
clear 
 that $\Bbb E^8$ equipped with the above complex structures and the 
flat metric is an OKT manifold.

To find non-trivial examples of eight-dimensional OKT geometries, 
we use the ansatz
$$
\eqalign{
ds^2&=H\, ds^2(\Bbb E^8)
\cr
c_{\mu\nu\rho}&=\Omega_{\mu\nu\rho}{}^\lambda \partial_\lambda H\ ,}
\eqn\esix
$$
where $\mu,\nu,\rho,\lambda=0,\dots, 7$ , $H$ is a function on
 $\Bbb E^8$ and $\Omega$ is a
four-form on ${\Bbb E}^8$ (We have raised the index of $\Omega$ with 
 the flat metric on $\Bbb
E^8$).  The choice of a conformally-flat metric in the ansatz is 
motivated by the  form of the moduli metric of the
$n=8$, $a=1$ black holes. In addition, we choose as complex structures
 those of \efive. 
The integrability of $I_a$ follows immediately because they   are
  constant tensors and so all
Nijenhuis conditions are  satisfied. The metric in \esix\ is clearly
 hermitian with respect to all
complex structures. So it remains to solve the last two conditions in \cfive.
The second condition in \cfive\ implies that
$$
\eqalign{
\Omega_{0abc}&=\varphi_{abc}
\cr
\Omega_{abcd}&=-{}^*\varphi_{abcd} \ . }
\eqn\eseven
$$
Therefore $\Omega$ is the ${\it Spin}(7)$ invariant  anti-self-dual 4-form
 in $\Bbb E^8$ associated
with the above octonionic structure. 
After some computation, the third condition in \cfive\ implies that
$$
\delta^{\mu \nu}\partial_\mu \partial_\nu H=0\ .
\eqn\eeight
$$
Hence, $H$ is a harmonic function on $\Bbb E^8$.  For this
computation, 
we have used the
following identities for the octonionic structure constants:
$$
\eqalign{
\varphi_{[ab }{}^{(f} \varphi_{c]d}{}^{e)}&={4\over 3}
 {}^*\varphi_{[abc}{}^{(f} \delta_{d]}^{e)}
-{1\over 3} {}^*\varphi_{abcd} \delta^{fe}
\cr
\varphi_{p[a }{}^{(f} {}^*\varphi_{bcd]}{}^{e)}&=-
 \delta_{p[a}\varphi_{bcd]} \delta^{fe}+
3 \delta_{p[a}\varphi_{bc}{}^{(f} \delta^{e)}_{d]}- 
 \varphi_{[abc} \delta_{d]}^{(f}
\delta^{e)}_p\ . }
\eqn\enine
$$
Note that $c$ is {\sl not} a closed three form.

Apart from the flat OKT structure in $\Bbb E^8$ for which $H=1$, the simplest
non-trivial OKT structure arises when
$$
H={1\over  |{\bf x}|^6}\ .
\eqn\eten
$$
Setting
$$
\rho={1\over 2 |{\bf x}|^2}\ ,
\eqn\eeleven
$$
the metric becomes
$$
ds^2=d\rho^2+4 \rho^2 d\Omega^2_{(7)}
\eqn\etwelve
$$
where $d\Omega^2_{(7)}$ is the standard round metric on $S^7$.  Geometrically,
the metric is a cone over the round 7-sphere of radius 2.  The metric
\etwelve\ is defined on ${\Bbb O}^*={\Bbb E}^8\\
\{0\}$ and is complete as $\rho\rightarrow \infty$, i.e as
 $|{\bf x}|\rightarrow 0$,  but 
has a conical singularity at finite distance as
$\rho\rightarrow 0$, {\it i.e.}\ $|{\bf x}|\rightarrow \infty$.
 To cure this problem, we set
$$
H=1+{1\over  |{\bf x}|^6}\ .
\eqn\ethirteen
$$ 
and we get a complete metric on ${\Bbb O}^*$ which interpolates 
between the flat metric near
$|{\bf x}|\rightarrow 0$ and the conical metric near
 $|{\bf x}|\rightarrow \infty$. 

Note that by taking the octonionic conjugate of the above
construction,
 we obtain another OKT
structure.  Each of these structures is invariant under the freely
 acting involution
$$
{\bf x}\rightarrow -{\bf x}\ .
\eqn\efourteen
$$

\section {{\bf Octonionic Multi-Models}}

In order to obtain more complicated models, including the Shiraishi 
metric for the relative moduli of the $n=8$, $a=1$ black holes, we begin 
with the metric
$$
ds^2=\sum^v_{a=1} du_a\, d\bar u_a+\sum_{i=1}^e
 {do_i\, d{\bar o}_i\over  (o_i\, {\bar o}_i)^3}\ ,
\eqn\eaone
$$
on
$$
({\Bbb O})^v\times (\Bbb O^*)^e\ ,
\eqn\eatwo
$$
where $\Bbb O^*= \Bbb E^8\\ \{0\}$ and $u_a$, $o_i$ are octonions. 
This metric is an OKT metric because
it is a sum of OKT metrics. Using these we can obtain more complicated OKT
geometries by imposing
 suitable constraints. For example, let $v=e=1$. Dropping the indices
 on $u,o$, we impose
$$
u-o=0
\eqn\eathree
$$
and recognize the OKT metric on $\Bbb E^8\\ \{0\}$, corresponding to the
harmonic function
$$
H=1+{1\over  |{\bf x}|^6}\ .
\eqn\eafour
$$
This corresponds to the Shiraishi
metric on the relative  moduli space of two $n=8$, $a=1$ black holes.  
Note that the constraint restricts us to the fixed point set of the ${\Bbb
Z}_2$ action
$$
(u+o, u-o)\rightarrow (u+o, -u+o)\ .
\eqn\eafive
$$

The multi-centre OKT metric on ${\Bbb E}^8\\ \cup_{i=1}^k\{{\bf
a}_i\}$
 with centres
$\{{\bf a}_i; i=1, \dots, k\}$ can be found by choosing
$v=1$ and $e=k$,  and then by imposing the conditions
$$
u+a_i-o_i=0\ ,
\eqn\easix
$$
where $a_i$ are $k$ constant octonions.

Finally, in order to derive the Shiraishi metrics for the relative
moduli of $k$ $n=8$, $a=1$ black holes, 
we can take
$v=k$ and
$e={1\over2} k (k-1)$ and replace the index $i$ on $o_i$ by the
 compound index $ab$ with $0<a<b\leq
k$. The constraints are
$$
u_a-u_b-o_{ab}=0\ .
\eqn\easeven
$$
To be more precise, agreement with the Shiraishi metrics requires an
appropriate rescaling the coordinates  to introduce the various
 \lq\lq mass parameters" that appear
in the black hole moduli metrics.  This completes our demonstration
 that the moduli space of $n=8$, $a=1$
black holes admits N=8b supersymmetry.

\section {{\bf OKT geometry in various dimensions}}

To find more general examples of OKT geometries in $8k$ 
dimensions, $k>1$, from those of the
previous section, we first write the coordinates of
 $\{ x^M; M=1, \dots, 8k\}$ of $\Bbb E^{8k}$ as
$$
x^M=x^{\mu i}\ ; \qquad \mu=1, \dots 8 \ , \qquad i=1, \dots, k\ .
\eqn\ebonea
$$ 
Then we use the  ansatz
$$
\eqalign{
ds^2&= U_{ij} d{\bf x}^i\cdot d{\bf x}^j
\cr
c&={1\over 3!}\Omega_{\mu\nu\lambda}{}^\rho \partial_{i\rho} U_{jk} 
dx^{\mu i}\wedge dx^{\nu j}\wedge
dx^{\lambda k}\ ,}
\eqn\ebone
$$ 
where $U$ is a function of $\Bbb E^{8k}$. For $c$ to be a
three-form, the matrix $U$ must satisfy
$$
\eqalign{
U_{ij}&=U_{ji}
\cr
\partial_{\rho i} U_{jk}&= \partial_{\rho j} U_{ik} \ .}
\eqn\ebtwo
$$ 
To complete the ansatz, we choose as complex structures
$$
{\bf I}_a= 1\otimes I_a\ ,
\eqn\ebthree
$$ 
where $\{I_a; a=1,\dots, 7\}$ are the complex structures
 on $\Bbb E^8$ associated with the
octonions as in \efive.
After some computation, we find that the above ansatz satisfies the conditions
of an OKT geometry provided that
$$
\partial_i\cdot \partial_j U_{k l}=0\ ,
\eqn\ebfour
$$
where $(\cdot)$ denotes the inner product with respect to the
 Euclidean 8-metric.

It remains for us to find examples of such geometries.  For this let us write
$$
U_{ij}=U^\infty_{ij}+\Delta U_{ij}\ ,
\eqn\ebfive
$$
where $U^\infty_{ij}$ is a constant matrix which can be thought as the asymptotic value of
$U_{ij}$ as $|{\bf x}^i|$ goes to infinity.  Solving \ebtwo\ and 
\ebfour\ for $\Delta U_{ij}$, we
find that it is a linear combination of
$$
{p_i p_j\over  |p_k {\bf x}^k-{\bf a}|^6}\ ,
\eqn\ebsix
$$
for different choices of k-vectors $\{p_1,\dots, p_k\}$ and different 
choices of centres ${\bf a}$. 
Therefore the most general expression for $\Delta U_{ij}$ is
$$
\Delta U_{ij}=\sum_{\{p\}} \sum_{{\bf a}} 
{p_i p_j\over  |p_k {\bf x}^k-{\bf a}(\{p\})|^6}\ .
\eqn\ebseven
$$
It is clear that this OKT geometry is associated with the co-dimension eight  
 planes 
$$
p_k {\bf x}^k-{\bf a}=0\ ,
\eqn\ebeight
$$
in  $\Bbb E^{8k}$.
It appears that the metric \ebone\ with ${\rm det}\, U^\infty\not=0$
and
 coefficients given in
\ebseven\ is non-singular on the complement of these planes in
$\Bbb E^{8k}$ provided that the planes, if they intersect,  intersect
 only pairwise.

\chapter{ Moduli Space Geometries for Black Holes  from Intersecting branes}

It is well-known by now that the various black hole solutions of 
supergravities in various dimensions have a ten- or eleven-dimensional
interpretation as IIA, IIB or M-theory intersecting  branes [\gppkt]. This has
led to the better understanding of the black hole solutions preserving less
than half of the eleven-dimensional supersymmetry. As we shall explain, this
interpretation is also helpful to determine some of the structure of the
moduli spaces of $n=4$, $a=1$ and $n=8, a=1$ black holes. In particular, the
ten-dimensional interpretation that we shall present below allows us (i) to 
determine the underlying worldvolume Lorentz invariance  and  (ii)  to find
the appropriate one-dimensional supersymmetry multiplet of the effective
theory of the black holes. Both points require some explanation. To explain
the former, we remark that the dimensionality of the effective theory of
the ten-dimensional solution is determined by the residual Lorentz invariance
that remains  unbroken  by the solution.  If it turns out that the  Lorentz
invariance group of the ten-dimensional solution is that of two or more
dimensions, then the effective action will resemble that  of an extended
object, {\it i.e.}\ that of a string or a brane.  This implies that the
effective theory of the corresponding black holes is obtained by reducing the
effective theory of the extended object to one dimension.  However, as we have
seen in section 2, the supersymmetry multiplets in one dimension that can be
obtained as reductions of the two-dimensional ones have  a geometry that is
more restrictive than that of the generic one-dimensional supersymmetry
multiplets.  This results in stronger constraints on the geometry of the
moduli of the black hole solution. To explain the latter point, we note that 
there are  
 supersymmetry projection operators associated with the
ten-dimensional
 solution.  A
close investigation of these operators reveals the type of
one-dimensional
 supersymmetry multiplet which should be used to describe
the effective theory of the associated black holes.

Let us begin with the $n=4$, $a=1$ black hole solution.  The most
symmetric
 ten-dimensional lifting of
this solution is that of the solution of IIB supergravity having the
 interpretation of two 3-branes
intersecting on a string.  This solution is known to preserve $1/4$ of the
supersymmetry of IIB theory. The explicit form of the solution in the string
frame is
$$
\eqalign{
ds^2&= H_1^{-{1\over2}} H_2^{-{1\over2}} ds^2(\Bbb E^{(1,1)})+ H_1^{-{1\over2}} H_2^{{1\over2}}
ds^2(\Bbb E^2)
\cr &
+  H_1^{{1\over2}} H_2^{-{1\over2}} ds^2(\Bbb E^2)+H_1^{{1\over2}}
 H_2^{{1\over2}}
ds^2(\Bbb E^4)
\cr
G_5&=F_5+*F_5\ ,}
\eqn\fone
$$
where
$$
F_5=\omega_1({\Bbb E}^{(1,3)})\wedge dH_1^{-1}+\omega_2({\Bbb
E}^{(1,3)})
\wedge dH_2^{-1}\ ,
\eqn\fonef
$$
$H_1, H_2$ are two harmonic functions on $\Bbb E^4$ associated with
the two 3-branes, $\omega_1({\Bbb E}^{(1,3)})$ and $\omega_2({\Bbb
E}^{(1,3)})$ are the volume forms along the world-volume directions of
the two 3-branes and the Hodge duality operation is taken with respect
to the metric
\fone. If we reduce this solution along the relative transverse
directions, then we get a self-dual string solution in six dimensions
which is determined by two harmonic functions [\papad].  (For
$H_1=H_2$, we recover the self-dual string solution of [\duff].)  Now
if we further reduce along the string direction and set $H_1=H_2$, we
find the $n=4$, $a=1$ black hole solution of section 2.

To continue, let us suppose that the string lies in the directions
$0,1$, 
the first 3-brane
lies in the directions $0,1,2,3$ and the second 3-brane lies in the 
directions $0,1,4,5$.  The
supersymmetry projections associated to the solution \fone\ are
$$
\eqalign{
\Gamma_{0123}\eta^1&=\eta^2 
\cr
\Gamma_{0145}\eta^1&=\eta^2 \ , }
\eqn\ftwo
$$
where $\eta^1, \eta^2$ are Majorana-Weyl spinors and $\{\Gamma_a; 
a=0,\dots, 9\}$
are the ten-dimen\-sional Gamma-matrices. Since the solution has a 
two-dimensional Lorentz invariance
and preserves
$1/4$ of the IIB supersymmetry, the effective theory must be  
two-dimensional with eight
supersymmetry charges. However as we have seen there are at least 
two different two-dimensional
supersymmetry multiplets with eight supersymmetry charges, 
{\it i.e.}\ the (4,4) multiplet and the (8,0)
one.  An examination of the supersymmetry conditions \ftwo\  
reveals that the two-dimensional chirality
operator $\Gamma_{01}$ does not have a definite sign when acting 
on the Killing spinors $\eta$. 
 This leads to a non-chiral effective theory on the string with
(4,4) supersymmetry and therefore to a geometry on the target space with two
copies of a {\it strong} HKT structure.  The effective theory of
 the associated black hole is just
the reduction of the two-dimensional effective theory along the 
spatial world-sheet direction. This
of course leads to a one-dimensional effective theory based on 
the N=8a multiplet for which, as we have
seen, its geometry is entirely determined by that of the 
two-dimensional (4,4) multiplet.  It is
remarkable that this is exactly what we have found by studying 
the Shiraishi metric on the moduli
space of
$n=4$, $a=1$ black holes.

Next let us turn to examine the $n=8$, $a=1$ black hole case. 
The most symmetric  description   of
this black hole solution in ten
dimensions is the IIA supergravity solution with the 
interpretation of a wave on a
string. The explicit ten-dimensional solution in the string frame is
$$
\eqalign{
ds^2&= H^{-1}_1 (du dv+ (H_2-1) du^2)+ ds^2(\Bbb E^8)
\cr
G_3&=dt\wedge dy\wedge dH_1^{-1}
\cr
e^{2\Phi}&=H_1^{-1}\ , }
\eqn\fthree
$$
where $u=t+y$ and $v=-t+y$, and $H_1$, $H_2$ are the harmonic functions on
$\Bbb E^8$ associated with the string and  the pp-wave, respectively. This
solution preserves $1/4$ of the IIA supersymmetry and reduces along the string
direction $y$ to a nine-dimensional black-hole solution with two harmonic
functions
$H_1, H_2$. If we next set
$H_1=H_2$, then we recover the $n=8$, $a=1$ black hole solution of section 2.

As in the previous case, let us consider the supersymmetry projections
associated with the wave-on-a-string solution of IIA supergravity above. 
Assuming that the string lies in the directions
$0,1$, we find
$$
\eqalign{
\Gamma_{01}\Gamma_{11}\eta&=\eta
\cr
\Gamma_{01}\eta&=\eta\ ,}
\eqn\ffour
$$
where $\eta$ is a ten-dimensional Majorana spinor. The 
solution \fthree\ has two commuting Killing
vectors along the the $u,v$ directions but no two-dimensional 
Lorentz invariance.  From this, we
conclude that the effective theory is best described by a
one-dimensional
 sigma model. Since the
ten-dimensional solution preserves $1/4$ of the supersymmetry, the
associated effective theory must have N=8 one-dimensional supersymmetry.  So
it remains to be determined whether the appropriate multiplet is the N=8a or
the N=8b. One way to find out which of the two multiplets is the relevant one 
is to observe that, from the string perspective, the supersymmetry preserved is
chiral, because in
\ffour\ the two-dimensional chirality operator $\Gamma_{01}$ acts on $\eta$
with a definite sign. But as we have explained in section 3, the
one-dimensional multiplet that is associated with chiral two-dimensional
supersymmetry with eight supercharges is N=8b.  It is remarkable that this
is exactly what we have found by analysing the Shiraishi geometry of the
moduli space of $n=8$, $a=1$ black holes.

\chapter{ Brane probes and  black holes}

The occurrence of non-trivial metrics on the moduli spaces for multiple
$p$-branes may easily be seen using ``test-brane'' probes, {\it i.e.}\ actions
for
$p$-branes situated in backgrounds corresponding to other $p$-branes, which may
be considered to be ``heavy'' and for which one may ignore, in leading order,
the back-reaction of the probe on the background metric. In all the cases that
we consider here, there is a ``no-force'' phenomenon for such probes:
they see no potential arising from the background, as a result of a
cancellation of forces due to gravity and scalars versus those from the
antisymmetric tensor fields. The general action for a $p$-brane probe with
tension
$T_\alpha$ and charge $Q_\alpha$ may be written
$$
I_{\rm probe} = -T_\alpha\int d^{p+1}\xi(-\det \partial_\mu x^m\partial_\nu
x^ng_{mn})^{1\over2}e^{{1\over2}\varsigma^{\rm pr}\vec
a_\alpha\cdot\vec\phi} + Q_\alpha\int\tilde A^\alpha_{[p+1]} ,\eqn\probeaction
$$
where 
$$\tilde A^\alpha_{[p+1]} = (p+1)^{-1}\partial_{\mu_1}x^{m_1}\cdots
\partial_{\mu_{p+1}}x^{m_{p+1}}A^\alpha_{m_1\cdots
m_{p+1}}d\xi^{\mu_1}\wedge\cdots d\xi^{\mu_{p+1}}
\eqn\probeb
$$ 
is the pull-back to the worldvolume of the gauge potential $A^\alpha$
that couples to the probe brane, $\vec a_\alpha$ are the ``dilaton
vector'' coupling parameters for the dilatonic scalars $\vec\phi$
occurring in the kinetic-term exponential prefactor for
$F^\alpha=dA^\alpha_{[p+1]}$, and $\varsigma^{\rm pr}=\pm 1$ according
to whether the probe is of electric or magnetic type.

     Three cases will suffice to illustrate the general phenomenon and
make contact with the discussions of the preceeding sections. In nine
spacetime dimensions, one has an electrically-charged black hole
solution, in whose background one may place an electrically-charged
probe particle, which however is coupled to a {\it different} field
strength ({\it i.e.}\ orthogonal to that of the background in charge
space). Despite the fact that the background and the probe couple to
different field strengths $F_\alpha$, a no-force condition is
nevertheless obtained [\boundstates, \rham]. For a probe coupling to a
field strength $F^\alpha$ orthogonal to that excited in the
background, the second term in
\probeaction  vanishes, and so the ``potential'' felt by the probe arises only
from the first term. The dilatonic factor in \probeaction  is
$\exp\{\ft12(a_{\rm probe}\cdot a_{\rm back}/|a_{back}|)\phi_{\rm
back}\}$.  Specifically, in the $N=2$ nine-dimensional maximal
supergravity theory, the dot product of the dilaton vectors
corresponding to the background and to the probe is $a_{\rm
probe}\cdot a_{\rm back}=-\ft{12}7$, while the diagonal dot products
are $a_\alpha\cdot a_\alpha=\ft{16}7$. For the electrically-charged
black hole solution, the nine-dimensional metric $ds^2=e^{2A}dx^\mu
dx^\nu\eta_{\mu\nu}+e^{2B}dy^mdy^m$ has $e^A=H^{-\ft37}$, while the
background dilatonic scalar is given by $e^{\phi_{\rm
back}}=H^{-2\over\sqrt7}$, where $H$ is the harmonic function
governing the solution. Putting together the metric and dilatonic
factors, one finds a potential $V_{\rm
probe}=e^Ae^{-{3\over2\sqrt7}\phi_{\rm back}}=1$, demonstrating the
zero-force condition for this probe-background
configuration. Continuing on to the next order in the velocities, one
finds a kinetic term for the moduli $-\ft12 T_\alpha
e^{-2A}e^{2B}\partial^\mu y^m\partial_\mu y^m$, giving a moduli-space
metric $H\delta_{mn}$. Similar analysis shows that the same result is
found in the two other test-brane cases related to our earlier discussions: a
nine-dimensional magnetic 5-brane in the background of another
magnetic 5-brane, again corresponding to orthogonal field
strengths, and similarly a six-dimensional electrically-charged string
probe in a magnetic string background, once again corresponding to
orthogonal field strengths. In all three of these cases, a constant
potential is obtained, and in all three cases the moduli-space metric
is given directly by the background's harmonic function:
$H\delta_{mn}$.

     In order to compare these brane-probe results with the exact
metrics given in \athree  (noting that all three cases correspond to $a^2=1$), 
one should separate the center-of-mass and relative moduli in \athree, giving 
a relative moduli metric for the two-center case
$$
ds_{\rm rel}^2 = 
{\pi^{{n\over 2}-1} \over 4( n-2) \Gamma({n\over 2}) }
\left[{\mu_1\mu_2\over \mu_1+\mu_2}+ {\mu_1\mu_2\over n-2}{1\over 
|{\bf x}_1-{\bf x}_2|^{n-2}}\right]|d({\bf x}_1-{\bf x}_2)|^2\ .
\eqn\relmetric
$$
Taking the limit $\mu_1/\mu_2\rightarrow\infty$ and dropping an overall factor
${\pi^{{n\over 2}-1} \over 4( n-2) \Gamma({n\over 2}) }\mu_2$, one obtains 
$H |d({\bf x}_1-{\bf x}_2)|^2$, with 
$$
H=1+{\mu_1\over
(n-2)|{\bf x}_1-{\bf x}_2|^{n-2}}\ ,\eqn\hval
$$
in agreement with the brane-probe results.

     The above test-brane discussion ignores back reaction
effects of the probe on the underlying spacetime, so the result for the moduli
space metric is not exact, {\it i.e.}\ it is only \lq\lq asymptotically" valid.
But for the three ``crossed field strength''  cases considered, this
discussion is enough to establish the existence of a non-trivial moduli-space
metric. This should be contrasted with the moduli-space metric for
similarly-oriented parallel 2-branes in eleven dimensions [\duffstelle] or the
analogous interaction between two parallel 1-branes (strings) in ten
dimensions [\dghrr]. In these latter cases, there is a zero-force condition
arising because of a cancellation between the second and first terms in
\probeaction. The moduli-space metric in these latter cases, however, proves
to be {\it flat}. This flat metric may be understood as a consequence
of the high degree of unbroken supersymmetry respected by the
probe-background system. For similarly-oriented parallel
eleven-dimensional 2-branes or ten-dimensional strings, the
probe-background system leaves unbroken a full $\ft12$ of the original
rigid supersymmetry of the theory, which on a two-dimensional
worldsheet corresponds to (8,8) supersymmetry. This degree of unbroken
supersymmetry is too restrictive to allow anything other than a flat
moduli-space geometry. The crossed-field-strength configurations,
however, leave unbroken only $\ft14$ of the supersymmetry,
corresponding either to worldsheet supersymmetry (8,0) (for the
nine-dimensional particles) or to (4,4) (for the nine-dimensional
5-branes or the six-dimensional strings).

     The test-brane analysis captures some of the general features of the
exact metrics that we have discussed in previous sections. In particular, the
moduli-space metric seen by a test-brane probe is of the same geometrical
class as the exact metric, although it represents only an asymptotic limit of
the exact metric. This follows the pattern of the analogous discussion for
moduli-space metrics for magnetic monopoles [\manton].

\section {{\bf Other five-dimensional black holes}}

Another well known example of a five-dimensional supersymmetric black
hole is that which arises in the computation of Bekenstein-Hawking
entropy by counting string states [\callanm].  This solution has the
ten-dimensional interpretation of a D-string within a D-5-brane and a
wave propagating along the D-string. The explicit solution in the
string frame is
$$
\eqalign{
ds^2&=H_1^{-{1\over2}} H_2^{-{1\over2}} \big( du dv+ (H_3-1)
du^2\big)+
 H_1^{{1\over2}}
H_2^{-{1\over2}} ds^2(\Bbb E^4)+ H_1^{{1\over2}} H_2^{{1\over2}} ds^2(\Bbb E^4)
\cr
G_3&=dt\wedge dy \wedge d H^{-1}_1+ \star dH_2
\cr
e^\Phi&= H^{-{1\over2}}_1 H^{{1\over2}}_2
}
\eqn\gone
$$
where $u=t+y$ and $v=-t+y$, the Hodge duality operation is taken with 
respect to the overall transverse
space $\Bbb E^4$, and
$H_1$,
$H_2$ and
$H_3$ are the harmonic functions associated to the string, 5-brane 
and pp-wave respectively. This
solutions preserves
$1/8$ of the supersymmetry of the IIB supergravity. Setting all the
 positions of the harmonic functions
to be the same and then compactifying the solution to five dimensions 
along all the worldvolume
directions of the D-5-brane, we find
 a black hole solution with finite  horizon area. In the following we 
shall take
$H_1$, $H_2$ and $H_3$ to have one centre.

Due to the interpretation of this  black hole solution  as a bound 
state of branes and a pp-wave,
one can compute the metric induced on one of the branes if it is considered as
a probe in the background generated by the remaining components. 
Since there are two different branes and a pp-wave in the bound state several
possibilities exist and the corresponding  probe metrics   were computed in
[\douglas, \douglaspolch].  In particular, if we consider D-5-brane probes in
the background generated by a wave on the string, the induced metric on the
D-5-brane is 
$$
ds^2= H_3 H_1\, ds^2(\Bbb E^4)\ .
\eqn\gtwo
$$
If $H_3=1$, then the theory, from the string perspective, has (4,4)
supersymmetry and the geometry on the moduli space is two commuting copies of
{\it strong} HKT structures.  This metric is the same as that on the relative
moduli space of two $n=4$, $a=1$ black holes. Therefore its multi-black hole
generalisations have already been discussed in the previous sections. The
most interesting case arises when both
$H_3$ and
$H_1$ are not constant.  In this case, from the two-dimensional
perspective,
 the
amount of supersymmetry preserved is (4,0).  This can be easily seen from an
analysis similar to that done for the black holes in the previous section. This
leads to a one-dimensional effective theory with
an N=4b supersymmetry multiplet.  As we have already mentioned, one of the
geometries compatible with this multiplet is the {\it weak} HKT geometry.  In
fact, it turns out that the relative moduli  space of two such black holes
admits a weak HKT structure.  The metric is as in \gtwo\ and the
torsion is
$$
c=3\, *d (H_3 H_1 )
\eqn\gthree
$$
where the Hodge operation has been taken with respect to the flat
 $\Bbb E^4$ metric.  The complex
structures are those of $\Bbb E^4$.

It is straightforward to construct generalisation of the above geometry similar
to that for the moduli metric of $k$ black holes.  For this we write the ansatz
$$
\eqalign{
ds^2&= U_{ij} d{\bf x}^i\cdot d{\bf x}^j
\cr
c&={1\over2}\,\epsilon_{\mu\nu\lambda}{}^\rho\, 
 \partial_{\rho k} U_{ij}\, dx^{\mu i}\wedge dx^{\nu
j}\wedge dx^{\lambda k}\ , }
\eqn\gfour
$$
where the coordinate is $\{x^{\mu i}; i=1,\dots, k; \mu=0,\dots,3\}$ 
and $U$ is a
function of $x$.  This geometry is a weak HKT provided that
$$
\eqalign{
U_{ij}&=U_{ji} 
\cr
\partial_{\mu i} U_{jk}&=\partial_{\mu j} U_{ik}\ ,}
\eqn\gfive
$$
with complex structures
$$
{\bf I}_r= 1\otimes I_r
\eqn\gsix
$$
where $I_r$ are three complex structures on $\Bbb E^4$.  A 
large class of asymptotically flat weak
HKT geometries can be obtained by choosing\foot{After this paper
appeared, the metric on the moduli space of these five-dimensional
black holes was obtained in [\modfive].} 
$$
U_{ij}=U_{ij}^\infty+ \Delta U_{ij}
\eqn\gseven
$$
where $U^\infty$ is the asymptotic value of $U$ as 
$|{\bf x}^k|\rightarrow \infty$,
$$
\Delta U_{ij}=\sum_{\{p\}} 
\sum_a p_i p_j f_{(\{p\})}(|p_k {\bf x}^k-{\bf a}(\{p\})|)
\eqn\geight
$$
and $f$ is any function that vanishes as $|{\bf x}^k|\rightarrow \infty$.

\chapter{Concluding Remarks} 

One motivation for obtaining the moduli space metrics
is to find the geodesics and hence to study the scattering of 
solitons in the low velocity limit. We shall not attempt to find
all the geodesics in all cases, but rather limit ourselves to making some
general remarks in the cases with $a=1$. The Lagrangian is given entirely by
the  kinetic energy $T$ which may be expressed as
$$
T=T_0 + T_{n-2},
\eqn\conone
$$   
where both  terms are positive and 
quadratic in velocities. The first is independent of the positions
${\bf x}_i$ while the second is homogeneous of degree $n-2$ in the position
variables.

The equations of motion are
$$
{d \over dt} {\bf p}_i= {\partial T_{n-2} \over \partial {\bf x}_i}\ ,
\eqn\contwo
$$
where ${\bf p}_i$ are the canonical momenta. Taking the dot product with 
${\bf x}_i$ and summing over $i$ gives
$$
\eqalign{
{d \over dt} \Bigl ( \sum {\bf x}_i \cdot{\bf p}_i \Bigr ) &
= 2T - (n-2) T _{n-2}
\cr
&= 2T_0 + (4-n) T_{n-2}\ ,}
\eqn\confour
$$
where we have used Euler's theorem on homogeneous functions.

Now assume that there are bound geodesics, {\it i.e.}\ geodesics that
are confined inside a compact set for all time. A special case would be 
closed geodesics. We can then average this equation over a large time to obtain a virial type relation:
$$
2\langle T_0 \rangle = (n-4) \langle T_{n-2} \rangle\ ,
\eqn\confive
$$
where $\langle ~~~\rangle$ denotes a time average. If $n\le 4$ we obtain an
immediate contradiction since the left hand side is positive while the right
hand side is negative or zero. Thus if $n=3$ or $n=4$ there can be
no bound geodesics for any number of solitons. By contrast if $n>4$ no
contradiction results, merely
a statement about the ratio of the two contributions to the energy.
Thus one might anticipate the existence of bound geodesics if $n>4$.
In the case of two solitons it is easy to see that there are (unstable)
 closed geodesics on the relative moduli space. 
This is because the asymptotically flat outer infinity 
(for which $|{\bf x}_1-{\bf x}_2| \rightarrow \infty$ is 
separated by a totally geodesic $(n-1)$-sphere 
from the asymptotically conical infinity for which 
$|{\bf x}_1-{\bf x}_2| \rightarrow 0$. 

For more than two solitons it seems rather likely that one could rigorously
establish the existence of closed geodesics using the Benci-Giannoni theorem
[\geod ] but we will not pursue this here. Geometrically it seems rather
plausible that every  closed geodesic will be unstable. 
One approach to  studying this might be to examine the curvature
of the metrics. 
It also seems rather 
likely that the geodesic motion and scattering
 will exhibit some of the chaotic
features  encountered in the closely related three-dimensional metrics
studied in [\cornish].

The classical moduli geometry of black holes may receive quantum 
corrections.  Some of these
corrections are due to the short-distance renormalisation effects of the
associated effective theory.  The cases that we have studied involve effective
theories with (4,4), (8,0) and (4,0) supersymmetries (in the two-dimensional
sense). The moduli geometries with (4,4) off-shell supersymmetry are protected
against quantum corrections because of the non-renormalisation theorem of
[\howepap]. In fact, since these moduli geometries have constant complex
structures, the Obata connection vanishes and the superfield constraints can
be solved exactly in terms of prepotentials allowing for a manifest
(4,4)-supersymmetric perturbation theory.  So the moduli metric of
$n=4$, $a=1$ black holes and some of the probe metrics of section 7 are 
expected to be exact in all
orders of perturbation theory. The same appears to apply for the moduli metric of
$n=8$, $a=1$ black holes. Finally, the probe metrics in section 7 with (4,0)
supersymmetry may receive corrections.  However these corrections 
are not due to
ultra-violet divergences but rather to finite local counterterms that are
necessary for the cancellation of anomalies in extended supersymmetry
transformations (for more details see [\howepapb]).

So far we have investigated the geometry on the moduli space of 
$n=4$, $a=1$ and $n=8$, $a=1$ black
holes. The geometry of moduli space of the rest of the $a=1$ black holes is
rather unclear.  It is likely that in the Shiraishi description of the moduli
space of all $4<n<8, a=1$ black holes there are missing moduli. This is in
direct analogy to  the missing moduli in the case of 
$n=3$, $a=1$ black holes [\manton] (see also section 2). However, it is worth
mentioning that all Shiraishi metrics on the moduli space of $a=1$
 black holes can be
constructed by taking the quotient of the moduli space of $n=8$, $a=1$ black
holes with a suitable group of translations.  To see this let us first
consider the case of the relative moduli,
${\cal M}_4^{(\ell)}$, of two $n=\ell$, $a=1$ black holes. It is clear that these
moduli can be identified with the quotient space
${\cal N}_4/{\Bbb R}^{8-\ell}$, where ${\cal N}_4$ is the relative 
moduli space of two $n=8$, $a=1$
black holes and ${\Bbb R}^{8-\ell}$ acts with translations on the 
first $8-\ell$ coordinates of
${\cal N}_4$.  More generally the moduli space of $k$ $n=\ell$, $a=1$ 
black holes, 
${\cal M}_{4k}^{(\ell)}$, can be identified with the quotient  
${\cal N}_{4k}/{\Bbb R}^{(8-\ell)k}$, where
${\cal N}_{4k}$ is the relative moduli space of $k$ $n=8$, $a=1$ black holes.

\vskip 1cm
\noindent{\bf Acknowledgments:}  We thank C. Bachas for helpful
comments. G.W.G and K.S.S. would like to thank the Isaac Newton Institute,
Cambridge, for hospitality.  The work of K.S.S. was supported in part
by the European Commission under TMR contract ERBFMRX-CT96-0045. G.P.
is supported by a University Research Fellowship from the Royal
Society.

\appendix

\section { Harmonic Forms}

The quantization of point particle mechanics may involve differential forms.
Special interest is attached to $L^2$ harmonic forms. For simplicity,
consider the relative moduli space of two solitons. By suitable rescalings
the metric may be brought to the form
$$
ds^2 = \bigl ( 1 + { 1\over r^{n-2}} \bigr ) 
( dr ^2 + r ^2 d \Omega _{n-1}^2 ),
$$
where $d \Omega _{n-1}^2$ is the standard round metric on $S^{n-1}$.
As noted in the previous section, if $n>4$ there is a totally geodesic
$(n-1)$-sphere located  at that value of $r$ for which $r^2 + 
{ 1\over r^{n-4}} $ is least. This suggest that if $n>4$ there is an 
associated harmonic form $(n-1)$-form equal to the 
volume form $\eta _{n-1}$ on $S^{n-1}$. 
As we shall now show, this is indeed true and  moreover the form
is in $L^2$. Obviously $\eta_{n-1}$ is closed:
$$
d \eta _{n-1}=0.
$$
The Hodge dual form is given by
$$
\star \eta _{n-1} = f(r) dr
$$
for some function $f(r)$ whose precise form we don't need. 
Evidently $\eta_{n-1}$ is co-closed and therefore harmonic.  
The $L^2$ norm of $\eta_{n-1}$
depends on the finiteness of the radial integral
$$ 
\int ^\infty _0 { dr \over r^{n-1}} (1 + 
{ 1\over r^{n-2} }) ^{-({ n\over 2} -1)}.
$$
The integral is convergent at infinity as long as $n>3$ For small
$r$ the integrand goes like $r^{ { 1\over 2} ( n^2 -6n+6)}$. Thus as long as
$n>4$, $\eta_{n-1}$ is indeed an $L^2$ harmonic form.

\section { Harmonic Spinors}

A similar analysis, but now with a negative result,
may be given for $L^2$ harmonic spinors. We use the conformal invariance
of the massless Dirac equation.
If $\psi_0$ is a solution of the massless Dirac equation in the flat metric
$$
ds = dr^2 + r^2 d \Omega ^2 _{n-1}
$$
then 
$$\psi = H^{-{n-1 \over 4}} \psi_0
$$
is a solution on the conformally-rescaled metric  
$$
ds = H \Bigl ( dr^2 + r^2 d \Omega ^2 _{n-1} \bigr ).
$$
The $L^2$ norm of $\psi$ depends on the radial integral:
$$
\int ^\infty _0 dr r^{n-1} H^{ 1\over 2} {\bar \psi _0} \psi _0.
$$
This becomes
$$
\int ^\infty _0 dr r^{n-1} ( 1+ { 1\over r ^{n-2} } ) ^{ 1\over 2} 
{\bar \psi _0} \psi _0.
$$ 
To obtain convergence at the upper end we must choose $\psi_0$ to decay
at large $r$. It follows that $\psi_0$ must be a linear combination of terms
of the form
$$
{ 1 \over r^{l+n-1} } \chi_l
$$
where $l=0,1,\dots $ and $\chi_l$ is an appropriate spinor harmonic 
on $S^{n-1}$ [\das]. The spinor field $\psi$ will be in $L^2$ if 
$$
{ 1\over r^{2l+2n-2}} r^{ n\over 2}
$$ is integrable as $r \rightarrow 0$. This requires:
$$
2l+ { 3 \over 2}n -2<1
$$
which can never be satisfied for any $n$. 
Given that one cannot find an $L^2$ harmonic spinor on the relative
moduli space of two solitons it seems rather unlikely that 
one can find one on the higher dimensional moduli spaces but we have
no general proof. Actually the previous result about the relative moduli space
of two solitons may be obtained
more simply by invoking Lichnerowicz's well known result
concerning harmonic spinors on spaces with non-negative Ricci scalar
in the cases  $n=4$ and $n=5$. The point is that since the function
$H$ 
is harmonic, the Ricci scalar $R$ of the
relative moduli space is given by
$$
R= - (n-1) { (n-6) \over 4} { (\nabla H)^2 \over H^3  }.
$$
Thus if $n=4$ or $n=5$ the Ricci scalar is positive and
 Lichnerowicz's argument applies. 

\refout
\end